\documentclass{aa}  

\usepackage{graphicx}
\usepackage[varg]{txfonts}
\newcommand{\Teff}{\ensuremath{T_{\mathrm{eff}}}}
\newcommand{\logg}{\ensuremath{\log g}}
\newcommand{\mass}{\ensuremath{\mathrm{M}_{\odot}}}

\newcommand{\FeH}{\ensuremath{\left[\mathrm{Fe}/\mathrm{H}\right]}}
\newcommand{\alphaFe}{\ensuremath{\left[\alpha/\mathrm{Fe}\right]}}

\newcommand{\GIRAFFE}{{\tt GIRAFFE}}
\newcommand{\UVES}{{\tt UVES}}
\newcommand{\COBOLD}{{\tt CO$^5$BOLD}}
\newcommand{\LHD}{{\tt LHD}}

\newcommand{\ATLAS}{{\tt ATLAS9}}

\newcommand{\SYNTHE}{{\tt SYNTHE}}

\newcommand{\WIDTH}{{\tt WIDTH9}}

\newcommand{\IRAF}{{\tt IRAF}}

\begin{document}

   \title{Sr}

   \title{Abundance of strontium in the Galactic globular cluster 47~Tuc\thanks{Based on observations obtained at the European Southern Observatory (ESO) Very Large Telescope (VLT) at Paranal Observatory, Chile.}}

\author{
	E.~Kolomiecas\inst{1}
	\and
	A.~Ku\v{c}inskas\inst{1}
	\and	
	J.~Klevas\inst{1}
	\and
	V.~Dobrovolskas\inst{1}
}

\institute{Institute of Theoretical Physics and Astronomy, Vilnius University,
	Saul\.{e}tekio al. 3, Vilnius, LT-10257, Lithuania\\
	\email{edgaras.kolomiecas@ff.vu.lt} }

\date{Received 11 September 2023; Accepted 24 November 2023}


\abstract
{}
{
We have determined Sr abundance in a sample of 31 red giant branch stars located in the Galactic globular cluster 47~Tuc with the aim to identify potential differences in the Sr abundance between first population (1P, Na-poor) and second population (2P, Na-rich) stars.
}
{
We derived the Na and Sr abundances from the archival spectra obtained with the \UVES\ spectrograph. To do this, we used 1D \ATLAS\ model atmospheres and a 1D local thermodynamic equilibrium spectral synthesis method. Particular attention was paid to assessing the potential impact of CN line blending on the obtained Sr abundances. Furthermore, we evaluated the potential influence of convection on the Sr line formation by using 3D hydrodynamical model atmospheres computed with the \COBOLD\ code.
}
{
Our results suggest a weak correlation between the abundances of Sr and Na. Together with a similar correlation between the abundances of Zr and Na determined in our previous study, our analysis of Sr suggests that polluters that have enriched 2P stars with light elements may have produced some s-process elements as well. The mean Sr abundance determined in 31 red giant branch stars of 47~Tuc is $\langle {\rm [Sr/Fe]} \rangle = 0.18\pm0.08$ (the error denotes the standard deviation due to the star-to-star abundance scatter). This value is within the range of the Sr abundance variation that is observed in Galactic field stars of similar metallicity. The mean [Sr/Zr] abundance ratio in our sample stars suggests that the two s-process elements could have been synthesized by either low-mass asymptotic giant branch stars ($M=1-4 {\rm M}_{\odot}$) or massive ($M=10-20 {\rm M}_{\odot}$) fast-rotating ($v_{\rm rot}=200-300$\,km\,s$^{-1}$) stars.
}
{}

\authorrunning{Kolomiecas et al.}
\titlerunning{Strontium in 47~Tuc}

\maketitle

%
\section{Introduction}

It is known that most if not all Galactic globular clusters (GGCs) consist of stars that belong to two (or more) stellar populations that differ in the abundances of the light chemical elements \citep[e.g.][]{BL18}. The chemical composition of the first population (hereafter 1P) is comparable to that of Galactic field stars at similar metallicity, whereas the second population (2P) is enriched in elements such as Li, N, Na, and Al and is depleted in O and Mg \citep[e.g.][]{C16,BL18}. While the origin of these populations is still unclear, one of the working hypotheses is that some of the 1P stars have produced sizeable amounts of light elements and then dispersed them into the interstellar medium, from which the 2P, now enriched, stars have formed. The notable candidates for these, the so-called polluters, include asymptotic giant branch stars \citep[AGBs; e.g.][]{DAVD16}, fast-rotating massive stars \citep[FRMS; e.g.][]{DCS07}, and supermassive stars \citep[$\sim10^4\,{\rm M}_{\odot}$, SMS; e.g.][]{DH14,GCK18}.

The situation is more ambiguous for the 1P--2P differences between the abundances of heavier elements, such as those produced during the s-processes. In theory, s-process elements can be synthesized in a number of sites (e.g., AGB stars and massive helium-burning stars; e.g., \citealt{CSP15,LC18}). However, while the light s-process elements (e.g., Sr, Y, and Zr, the so-called first s-process peak) may be produced in both AGB and massive stars, heavier s-process elements (e.g., Ba and La, the second s-process peak) are expected to originate predominantly from the lower-mass AGB stars \citep{C18}. The ratio of light-to-heavy s-process elements might therefore help to identify the candidate polluters in the GGCs. 

Observationally, the 1P--2P differences in s-process elements are seen in the most massive ($> 10^6$~M$_{\odot}$) GGCs that belong to a relatively small group of the so-called anomalous or Type~II GGCs. These GGCs  also show significant spreads in the abundances of Fe and Fe-group elements \citep{MMK15,MMR19}. 
In contrast, 1P--2P abundance differences as well as \mbox{(anti-)correlations} between the abundances of light and s-process elements have only been detected in a few Type~I GGCs, even though clusters of this type constitute the majority of the GGCs.

For example, in their study of 106 red horizontal branch (RHB) stars in 47~Tuc, \citet{GLS13} detected a weak but statistically significant Ba--Na correlation. However, this result has not been corroborated by subsequent investigations of 47~Tuc by \citet{TSA14} and \citet{DKK21}, the latter study was based on the analysis of Ba abundance in 261 red giant branch (RGB) stars. Moreover, no 1P--2P difference has been detected in the abundance of La, another second s-process peak element \citep[][123 RGB stars]{CPJ14}. 
However, \citet{FBB2021,FVG2022} have recently suggested Ce--N and Ce--Al correlations in  NGC~6380, Tonantzintla~2, and, possibly, 47~Tuc. Because Ce, Ba, and La all belong to the second s-process peak, it is expected that these elements would follow the same abundance patterns. It is therefore puzzling that at least in 47~Tuc, the currently available data on Ce do not suggest this to be the case, which clearly warrants further analysis of Ce abundances in a larger sample of cluster stars. 

As for the light s-process elements, a possible existence of the 1P--2P differences and/or (anti-)correlations with the abundances of light chemical elements (O, Na) has been suspected in several studies \citep[e.g.][]{VG11,DOCL13,SSG16}.
More recently, our analysis of the Zr abundance in 237 RGB stars in 47~Tuc, the largest stellar sample in which this element has been investigated in any GGC so far, has revealed a weak Zr--Na correlation \citep{KDK22}. This raises the question whether similar correlations may be seen for other light s-process peak elements as well, such as Rb, Sr, and Y, in this or other GGCs.

In this paper, we study the abundance of Sr in the RGB stars in 47~Tuc. Because both Sr and Zr belong to the first s-process peak, it is to be anticipated that both elements would show similar 1P--2P abundance differences and (anti-)correlations with the abundances of light chemical elements. The correlation between Sr and Na would be strong evidence that the polluters had to produce light s-process elements (Zr, Sr), in addition to the light elements (e.g., Na). The only investigation of the Sr abundance in 47~Tuc carried out so far was based on the analysis of eight subgiant branch (SGB) and three main-sequence turn-off (TO) stars \citep{J04}. The analysis revealed neither a significant Sr abundance spread nor correlations with the abundances of light elements. This might at least in part be caused by the small sample of stars studied, especially given the small spreads and/or the weak correlations expected for the s-process elements 
\citep[cf.][]{FBB2021,FVG2022,KDK22}. The goals of the present work therefore were (a) to determine Sr abundances in a larger sample of stars, and (b) to check whether any 1P--2P abundance differences and/or correlations may exist in case of Sr in this GGC.

\section{Observational data and atmospheric parameters\label{sect:obs_data}}

\subsection{Stellar sample and observed spectra\label{sect:obs_spectra}}

The abundance analysis of Fe, Na, and Sr was performed using archival spectra obtained during four observational programs with the \UVES\ spectrograph mounted on the VLT UT2 telescope (programs 072.D-0777(A), PI: Francois; 073.D-0211(A), PI: Carretta; 084.B-0810(A), 086.B-0237(A), PI: Sbordone;  088.D-0026(A), PI: McDonald). To the best of our knowledge, these are the only spectra of 47~Tuc that are available from the ESO Advanced Data Products (ADP) archive\footnote{\url{http://archive.eso.org/wdb/wdb/adp/phase3_spectral/form}} in which at least one Sr line could be measured reliably (we used the 650.3991\,nm \ion{Sr}{i} line for the abundance analysis; Sect.~\ref{sect:abund-sr}; see also Appendix~\ref{app-sect:arch_spectra} for more details of the spectrum selection).
The observing log is provided in Table~\ref{tab:obs-journal}. All target spectra were continuum-normalized using the \texttt{splot} task in the \IRAF\ package \citep{T86}. The typical signal-to-noise ratio was $\sim90-160$ at 640\,nm. The radial velocities were determined by using the \texttt{fxcor} task under \IRAF. The cross correlation with the \texttt{fxcor} task was made using a template synthetic spectrum, which was computed using average atmospheric parameters of the studied stars ($\Teff=$ 4155\,K, $\logg=$ 1.18).

\begin{table}[b]
	\begin{center}
		\caption{Spectroscopic data.
			\label{tab:obs-journal}}
		\resizebox{\hsize}{!}{%
			\begin{tabular}{ccccccc}
				\hline\hline
				\noalign{\smallskip}
				Programme       & Date of      & $\lambda_{\mathrm{interval}}$, & $R$  & Exposure, & Number of \\
			                	& observation  &    nm                         &    & s         & targets     \\
				\hline\noalign{\smallskip}
				072.D-0777(A)   & 2003-10-20   & 472 -- 683   & $\sim$50500     & 3000--8600    & 10       \\
				073.D-0211(A)   & 2004-06-26   & 472 -- 683   & $\sim$50500     & 1600--4315    & 11       \\
				084.B-0810(A)   & 2009-11-16   & 472 -- 683   & 107200          & 3600         & 5        \\
				086.B-0237(A)   & 2010-10-03   & 472 -- 683   & 107200          & 3600         & 8        \\
				088.D-0026(A)   & 2011-11-26   & 472 -- 683   & $\sim$48000     & 4080--8151    & 4       \\
				\hline
		\end{tabular}}
	\end{center}
\end{table}

\begin{figure}[tp]
	\centering
	\includegraphics[width=7.5cm]{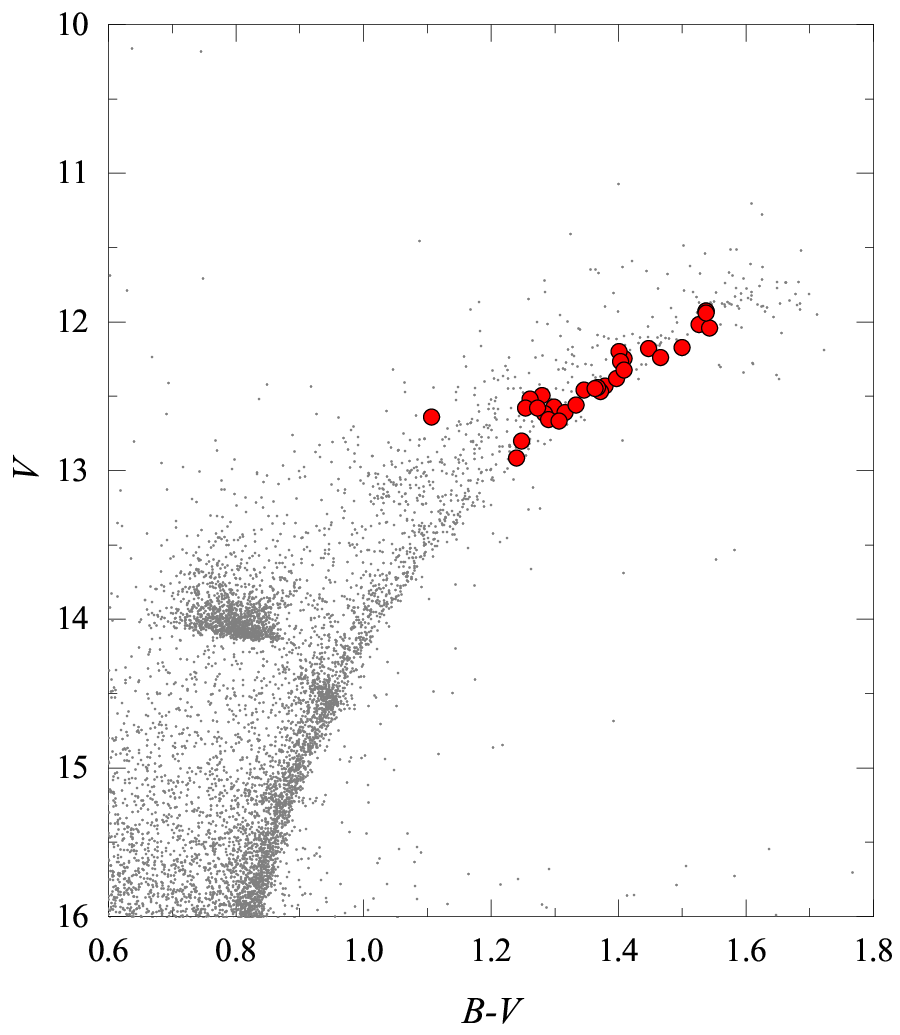}
	\caption{Color-magnitude diagram of 47~Tuc with the target RGB stars marked as red circles (photometry from \citealt{BS09}). 
	}
	\label{app-fig:CMD}
\end{figure}

We furthermore removed six stars from the 072.D-0777(A) sample with poor-quality spectra. These six stars also had the highest effective temperatures (\Teff $>$ 4500\,K) and the weakest \ion{Sr}{i} lines. One more star was removed from this sample because it had a very high effective temperature (\Teff\ = 6142\,K) and the \ion{Sr}{i} line was too weak for an abundance determination. The final sample of stars we used consisted of 31 RGB stars (Fig.~\ref{app-fig:CMD}). It contains 17 stars in common with the sample used in \citet{KDK22} for the Zr abundance determination (no Sr abundances were obtained in the latter study because of the lack of Sr lines in their \GIRAFFE\ spectra). By using their full spatial velocities (radial velocities plus proper motions from the \textit{Gaia} DR3 catalog; \citealt{GC23}), we verified that all sample stars are cluster members (Appendix~\ref{app-sect:clust_memb}).

\subsection{Atmospheric parameters\label{sect:atm_par}}

\begin{figure}[t]
	\centering
	\includegraphics[width=8.0cm]{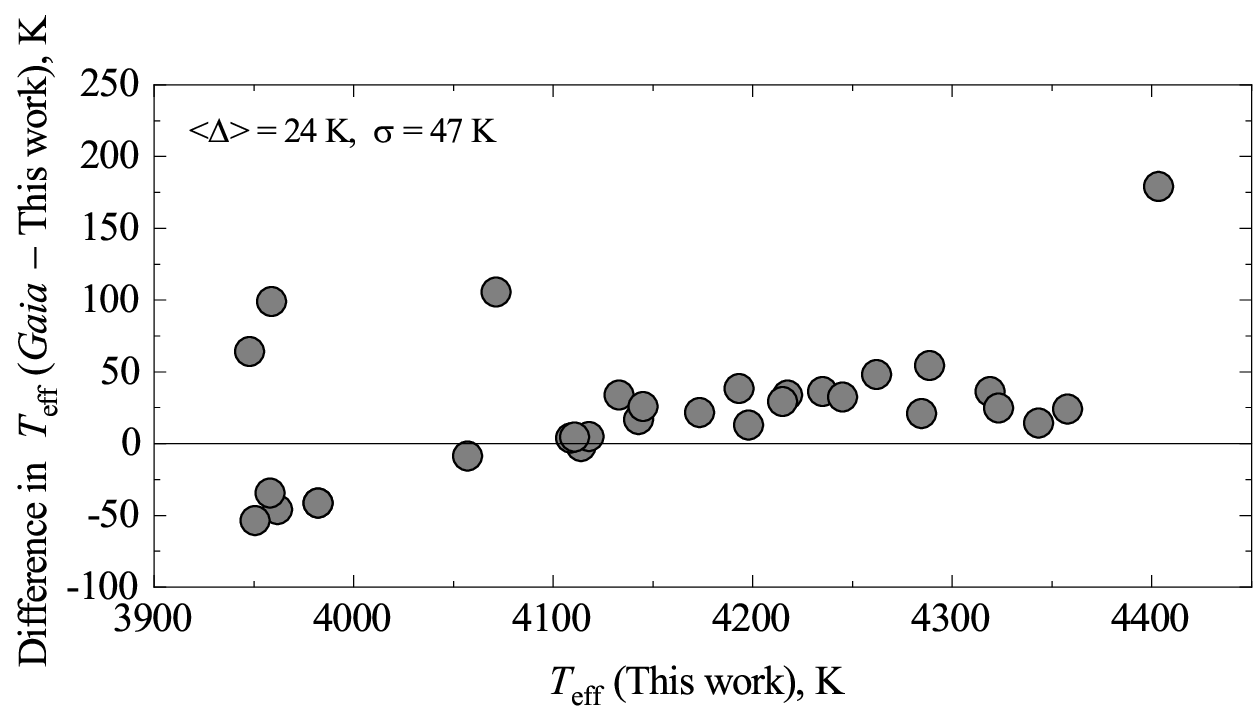}
	\caption{Effective temperatures determined using the \textit{Gaia}~DR3 photometry vs. those obtained using photometry from \citet[][the latter values of effective temperatures were used in this work]{BS09}. The mean difference between the two sets of effective temperatures and the scatter around this value (measured as the standard deviation, $\sigma$) are shown in the top left corner of the figure.}
	\label{fig:GAIA-Our_Teff}
\end{figure}

\begin{figure}[h]
	\centering
	\includegraphics[width=\hsize]{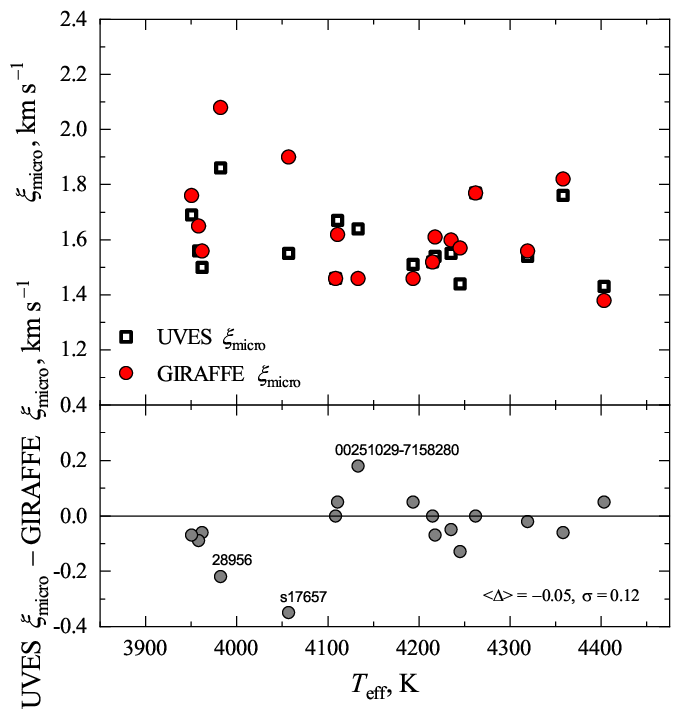}
	\caption{Difference between the microturbulence velocities determined from the \ion{Fe}{i} lines in the \UVES\ and \GIRAFFE\ spectra. The mean difference between the two sets of values and their scatter around the mean (measured as standard deviation, $\sigma$) are shown in the bottom panel of the figure.}
	\label{fig:Vmic-commonstars}
\end{figure}

The effective temperatures of the target stars were determined using the $\Teff-(V-I)$ calibration of \citet{RM05} and photometric measurements from \citet{BS09}. Although \textit{Gaia} DR3 photometry is available for the target stars, we used \citet{BS09} $VI$ data to maintain consistency with our previous studies of 47~Tuc (e.g., \citealt{CKK17, DKK21, KDK22}). We note that the average difference between the effective temperatures obtained with the \textit{Gaia}~DR3 photometry and the color-effective temperature calibration from \citet{MBM21} and those determined using the \citet{BS09} data and the $\Teff-(V-I)$ calibration of \citet{RM05} is small, $\sim$24\,K (Fig.~\ref{fig:GAIA-Our_Teff}).
We verified our obtained \Teff\ values by checking the trends in the plane of the iron abundance versus the line excitation potential, which in the majority of cases were close to zero.

The surface gravities, \logg, were derived using the classic relation between the effective temperature, stellar mass, luminosity, and surface gravity.
This was done for consistency with our previous works (e.g., \citealt{CKK17, DKK21, KDK22}). Furthermore, the mean iron abundance determined for the target stars by using the \ion{Fe}{i} and \ion{Fe}{ii} lines agrees to within 0.01\,dex, which indicates good agreement between the \logg\ values determined from photometry and those obtained via the ionization balance condition.
To do this, we assumed an identical stellar mass of 0.89\,\mass\ for all target RGB stars, as determined from the Yonsei-Yale isochrone\footnote{Version 2 from \url{http://www.astro.yale.edu/demarque/yyiso.html}} (12 Gyr, [M/H] = $-0.68$; \citealt{YY2}).
While some systematical differences may be inferred from the comparison of the Yonsei-Yale isochrone and the stellar parameters derived from observations, the differences fall within the scatter between the isochrones computed by different groups ($\pm$\,50\,K; e.g., \citealt{Dartmouth08}; \citealt{BaSTI21}). 

The microturbulence velocity, $\xi_{\rm micro}$, was determined individually for each target star by requiring that the \ion{Fe}{i} lines with different equivalent widths ($W$) would provide the same Fe abundance. We discarded strong \ion{Fe}{i} lines ($W$ > 16\,pm) because of their lower sensitivity to the changes in $\xi_{\rm micro}$. The list of the \ion{Fe}{I} lines used in the determination of Fe abundances, as well as the microturbulence velocities, was taken from \citet{KDK22}.

The microturbulence velocities obtained from the \ion{Fe}{i} lines in this work (\UVES\ spectra) and those in \citet[][\GIRAFFE\ spectra]{KDK22} typically agree to within 0.10\,km~s$^{-1}$ for the 17 stars common to the two samples (Fig.~\ref{fig:Vmic-commonstars}). The exception are 3 stars that are among the coolest stars in the sample and are affected by stronger line blending, which makes the continuum determination more difficult, and consequently, results in less accurate Fe abundances and microturbulence velocities.

The atmospheric parameters of the target RGB stars used in this study are provided in Table~\ref{app-tab:data-list}.

\section{Abundance analysis}

The abundances of Sr and Na were determined by applying the local thermodynamic equilibrium (LTE) spectral synthesis technique. For the abundances of Na, we further applied \mbox{non-local} thermodynamic equilibrium (NLTE) abundance corrections from our previous work \citep{KDK22}.
The Fe LTE abundances were obtained by using the equivalent width method. In all cases, we used 1D hydrostatic \ATLAS\ stellar model atmospheres \citep{K93,S05} with the $\alpha$-element (O, Ne, Mg, Si, S, Ar, Ca, and Ti) enhanced chemical composition, $\alphaFe=+0.4$. The atomic parameters of the spectral lines were taken from the VALD3 database \citep{RPK15} and are summarized in Table.~\ref{tab:line-list}.

\begin{table}
	\begin{center}
		\caption{Atomic parameters of the spectral lines used in this study.
			\label{tab:line-list}}
		\resizebox{\hsize}{!}{%
			\begin{tabular}{llcrllc}
				\hline\hline
				\noalign{\smallskip}
				Element & $\lambda$, nm  & $\chi$, eV & log \textit{gf} & log $\gamma_\mathrm{rad}$ & log $\frac{\gamma_4}{N_\mathrm{e}}$ & log $\frac{\gamma_6}{N_\mathrm{H}}$ \\
				\hline\noalign{\smallskip}
				\ion{Na}{i}  & 615.4225 & 2.102 & $-1.547$ & 7.85                & $-4.39$     & $-7.28$               \\
				\ion{Na}{i}  & 616.0747 & 2.104 & $-1.246$ & 7.85                & $-4.39$     & $-7.28$               \\							
				\ion{Sr}{i}  & 650.3991 & 2.259 & $0.320$  & 7.69                & $-5.73$     & $-7.72$              \\						
				\ion{V}{i}	 & 650.4165 & 1.183 & $-1.280$ & 7.66                & $-6.06$     & $-7.76$                   \\
				\hline
		\end{tabular}}
	\end{center}
\end{table}

\subsection{Determination of the Fe abundances\label{sect:Fe-abund}}

\begin{figure}[tb]
	\centering
	\includegraphics[width=\hsize]{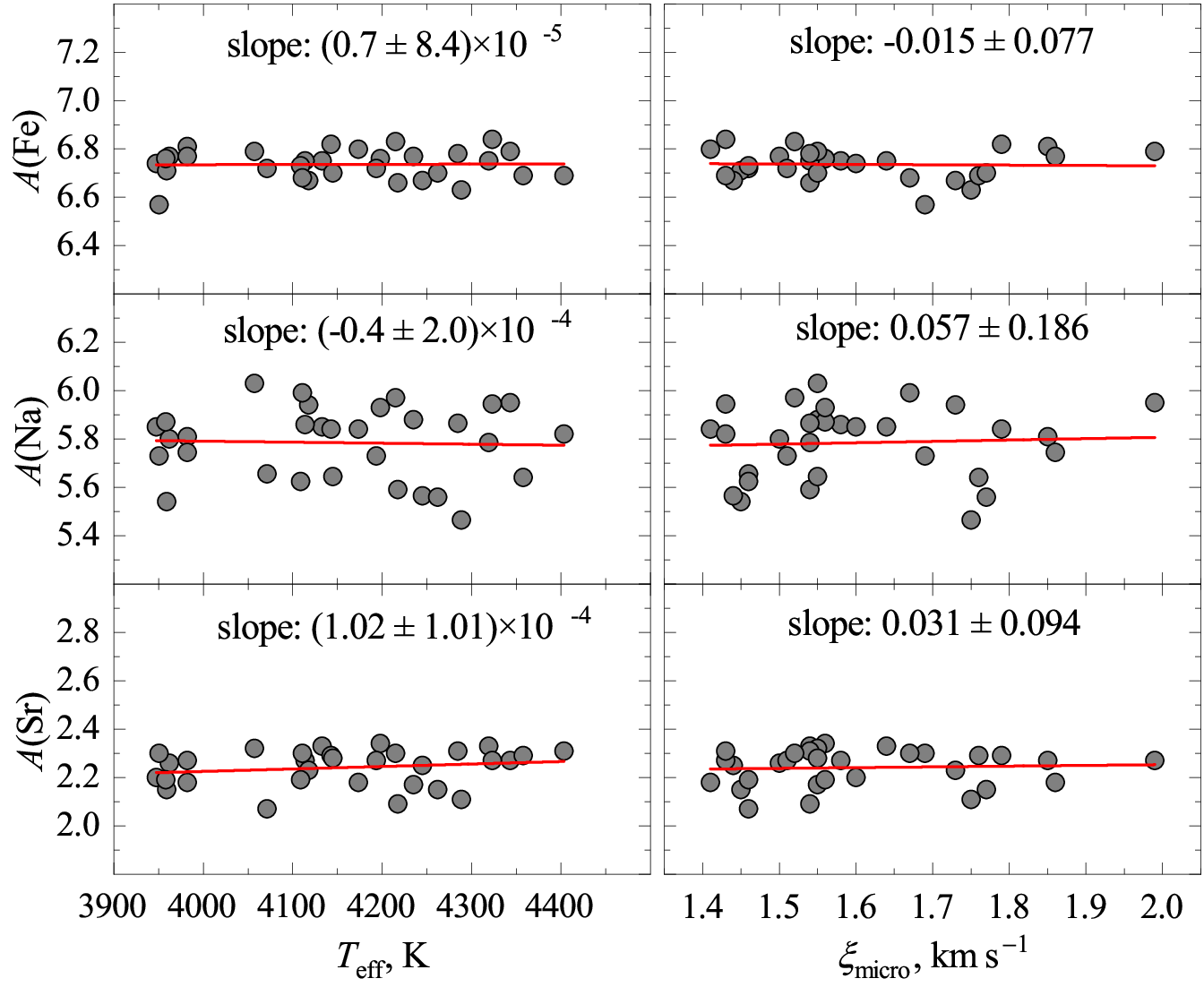}
	\caption{Abundances in the target RGB stars vs the effective temperature and microturbulence velocity in individual stars. The linear fits to the data are shown as red lines.}
	\label{fig:Teff_vmic_COMBO}
\end{figure}

\begin{figure}[t]
	\resizebox{\hsize}{!}{\includegraphics{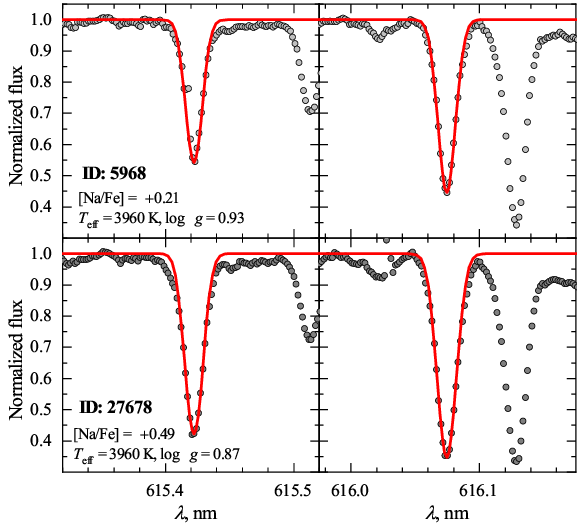}}
	\caption{Typical observed (dots) and best-fit synthetic LTE profiles of the \ion{Na}{i} 615.4225 and 616.0747\,nm lines (solid red lines) in the \UVES\ spectra of two RGB stars in 47~Tuc: Na-poor (5968, 1P, top row) and Na-rich (27678, 2P, bottom row). The identification numbers of each star, their atmospheric parameters, and the determined Na abundances are provided in the leftmost panels of each row.
	}
	\label{fig:spectral-lines}
\end{figure}

\begin{figure}[h]
	\centering
	\includegraphics[width=\hsize]{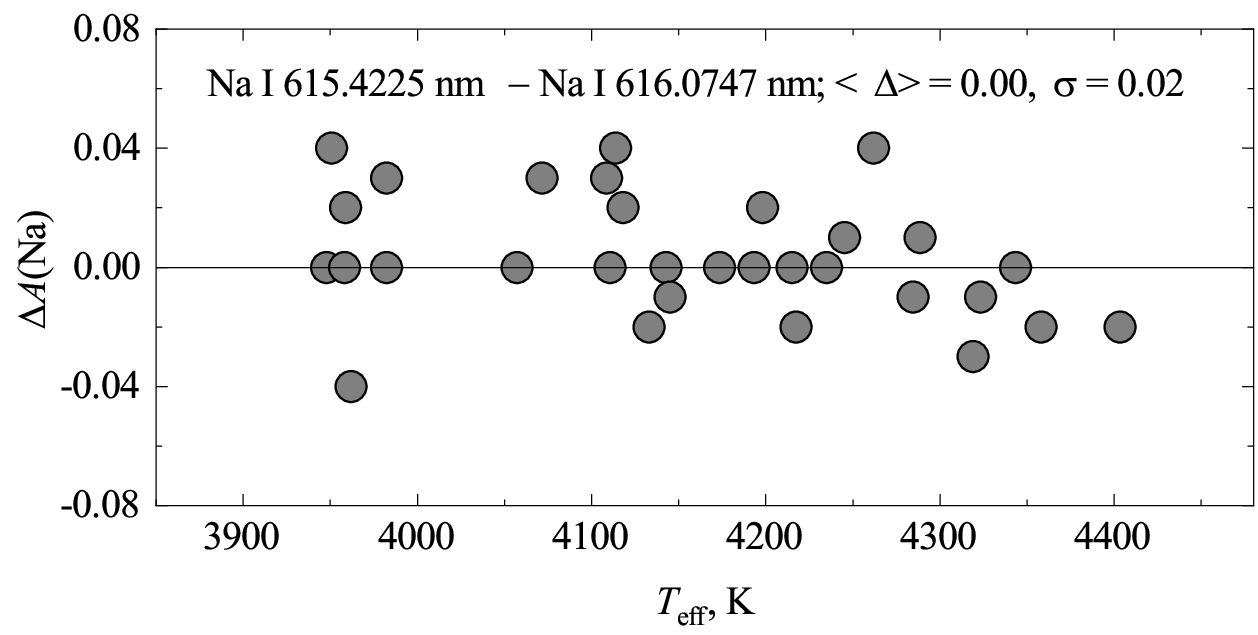}
	\caption{Difference in the Na abundance obtained in the target stars  from the two \ion{Na}{i} lines, plotted against the effective temperature. }
	\label{fig:Na-difference-Teff}
\end{figure}	

The 1D~LTE iron abundances were determined by fitting Gaussian profiles to the observed \ion{Fe}{i} lines using the \texttt{splot} task in \IRAF. The measured line equivalent widths, $W$, were used to obtain Fe abundances with the \WIDTH\ code \citep{S05} by using the 1D hydrostatic \ATLAS\ model atmospheres. We used 15--28 \ion{Fe}{i} spectral lines per star, with the lower level excitation potentials in the range of $\chi=2.18-4.61$\,eV, taken from the list used in \citet{KDK22}.
Lines with $\chi<2$\,eV were discarded to minimize the impact of non-local thermodynamic equilibrium (NLTE) effects. We found no dependence of the derived Fe abundances on \Teff\ and/or $\xi_{\rm micro}$ of the target stars (Fig.~\ref{fig:Teff_vmic_COMBO}).

The determined mean Fe abundance for the 31 target RGB stars, $\langle{\rm [Fe/H]}\rangle=-0.81\pm 0.06$ (the error denotes the standard deviation due to the star-to-star abundance scatter), agrees well with the Fe abundances in 47~Tuc obtained in other studies: $\langle{\rm [Fe/H]}\rangle=-0.74\pm0.05$ by \citealt[][114 RGB stars]{CBG09a}; $\langle{\rm [Fe/H]}\rangle=-0.77\pm0.08$ by \citealt[][44 RGB stars]{WPC17}. We used the solar iron abundance of ${\rm \emph{A}(Fe)_{\odot}}=7.55 \pm 0.06$ determined by \citet{KDK22}, where the same \ion{Fe}{i} line list was used as in this work. The average Fe abundance measurement error, $\sigma_\mathrm{i}(\mathrm{Fe})$, is $\sim$0.17\,dex (see Appendix~\ref{app-sect:abund-errors} for more details). The Fe abundances measured for each individual star are provided in Table~\ref{app-tab:data-list}.

\subsection{Determination of the Na abundances}

\begin{figure}[t]
	\centering
	\includegraphics[width=\hsize]{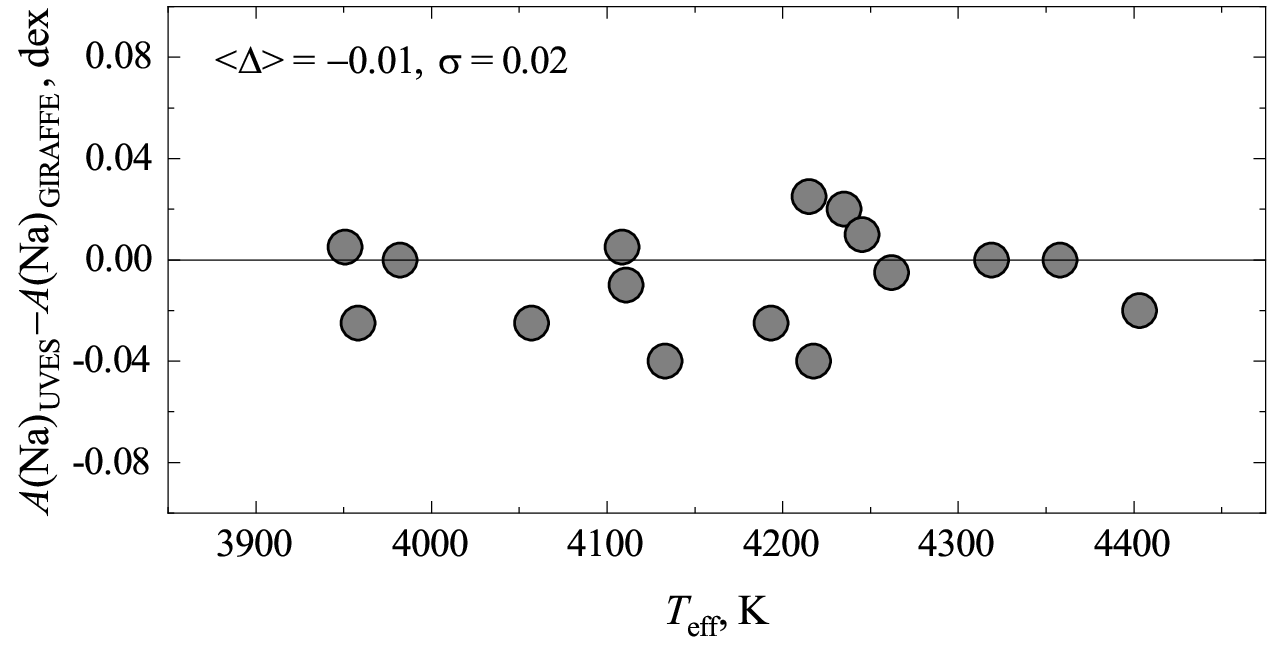}
	\caption{Difference in the Na abundance determined by using \UVES\ (this work) and \GIRAFFE\ spectra \citep{KDK22} for 16 stars in common between the two studies. The sodium abundance is an average obtained from the two \ion{Na}{i} lines.}
	\label{fig:Na-UVES-GIRAFFE}
\end{figure}	

\begin{figure}[t]
	\centering
	\includegraphics[width=7.0cm]{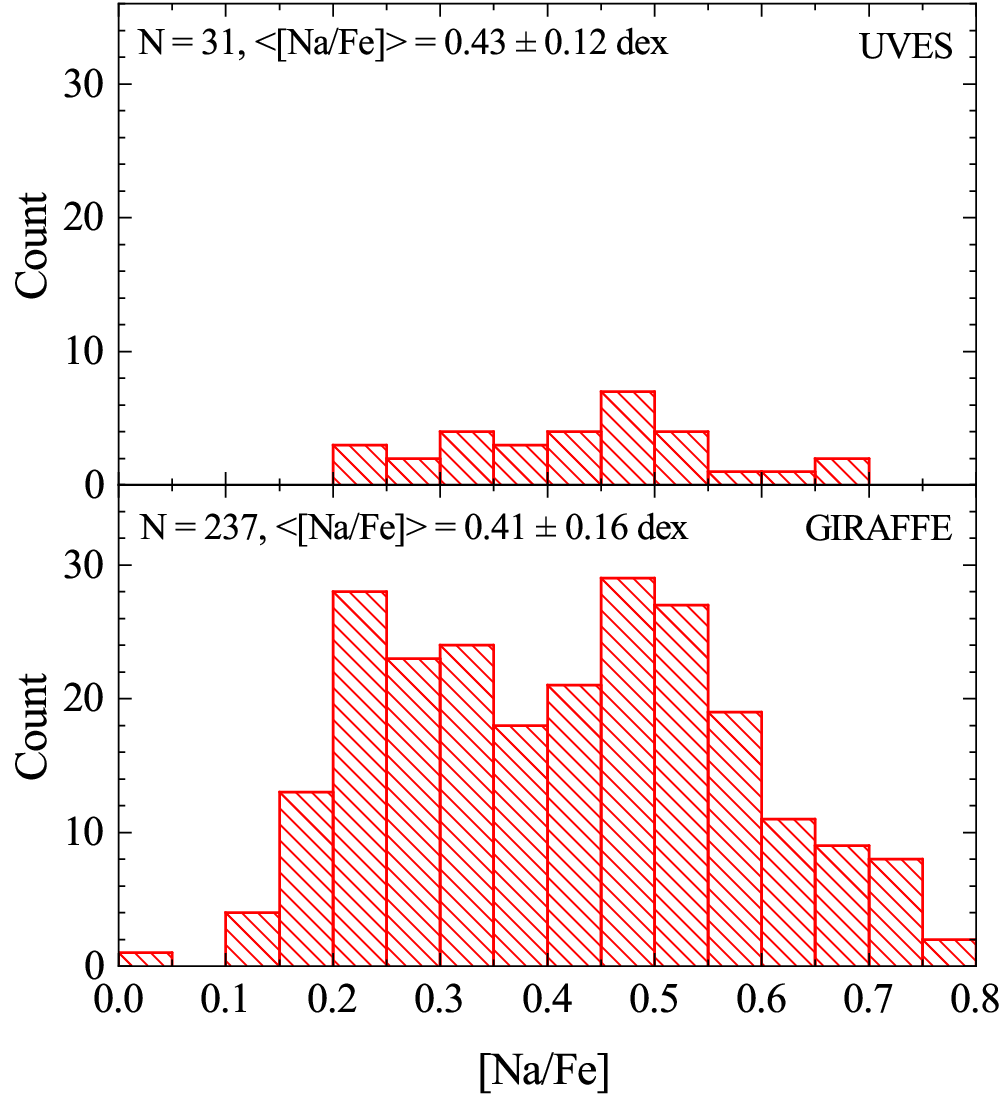}
	\caption{Histogram of the [Na/Fe] abundance ratios in stars of 47~Tuc analyzed in this work (top) and those studied in \citet[][bottom]{KDK22}. The number of stars in each sample is shown in the top left corner of each panel, together with the average [Na/Fe] abundance ratio and the standard deviation of the mean. }
	\label{fig:NaFe_hist}
\end{figure}

The 1D~LTE Na abundances were determined by applying the spectral synthesis method, with the \SYNTHE\ package used to compute the synthetic line profiles \citep{S05}. We used two \ion{Na}{i} lines for the abundance determinations, with their central wavelengths located at 615.4225 and 616.0747\,nm (Table~\ref{tab:line-list}). The abundances were obtained by fitting synthetic \ion{Na}{i} line profiles to those in the observed spectra. 
The obtained 1D~LTE Na abundances were further corrected for NLTE effects by applying an identical NLTE abundance correction equal to $-0.14$\,dex. The latter was obtained from our earlier study of Na NLTE abundances in the sample of 237 RGB stars in 47~Tuc \citep{KDK22}, whose atmospheric parameter range fully covers those of the target stars analyzed in the present work.

The solar Na abundance values, ${\rm \emph{A}(Na)_{\odot}^{\rm LTE}}=6.24$\,dex, ${\rm \emph{A}(Na)_{\odot}^{\rm NLTE}}=6.17$\,dex, were taken from \citet{KDK22}, who measured them using the same Na lines as we did. Typical fits of the synthetic profiles to the \ion{Na}{i} lines observed in the target star spectra are shown in Fig.~\ref{fig:spectral-lines}.

There is no dependence between the two-line-averaged Na abundance and \Teff\ and/or $\xi_{\rm micro}$ of the target stars (Fig.~\ref{fig:Teff_vmic_COMBO}). The Na abundances obtained from the two \ion{Na}{i} lines also agree well (Fig.~\ref{fig:Na-difference-Teff}). 

The mean [Na/Fe] ratio determined in 31 RGB stars is $\langle {\rm [Na/Fe]} \rangle = 0.43\pm0.12$ (the error denotes the standard deviation due to the star-to-star abundance scatter). This result agrees well with the Na abundance derived in 47~Tuc in our previous study, $\langle {\rm [Na/Fe]} \rangle = 0.41\pm0.16$ (\citealt{KDK22}, 237 RGB stars), and also in other studies, for instance, $\langle {\rm [Na/Fe]} \rangle = 0.47\pm0.15$ (\citealt{CBG09a}, 147 RGB stars), $\langle {\rm [Na/Fe]} \rangle = 0.36\pm0.18$ (\citealt{WPC17}, 27 RGB stars). The Na abundances obtained in this study (\UVES\ spectra) and that of \citet[][\GIRAFFE\ spectra]{KDK22} in 17 stars in common to the two studies also agree well (Fig.~\ref{fig:Na-UVES-GIRAFFE}). The average sodium abundance measurement error, $\sigma_\mathrm{i}(\mathrm{Na})$, is $\sim$0.10\,dex (see Appendix~\ref{app-sect:abund-errors} for more details). 

The range in the Na abundances obtained by us for the 31 target stars is slightly smaller than that determined by \citet[][Fig.~\ref{fig:NaFe_hist}]{KDK22}. This is to be expected, however, because the sample size used by \citet{KDK22} is nearly eight times larger than the one employed in this study. Nevertheless, the span in Na abundances in the target star can be deemed representative of the full sample. To assess this, we took the mean Na abundances determined from the \GIRAFFE\ and \UVES\ spectra and applied the two-sample $t$-test to determine statistical significance of the differences between these two samples (assuming the null hypothesis that the two samples are identical). The results of this analysis indicate that there is indeed no difference between the average Na abundances determined from \GIRAFFE\ and \UVES\ spectra ($p=0.56$).

The determined Na abundances are provided for each individual star in Table~\ref{app-tab:data-list}.

\subsection{Determination of the Sr abundances\label{sect:abund-sr}}

\begin{figure}[tb]
	\centering
	\includegraphics[width=\hsize]{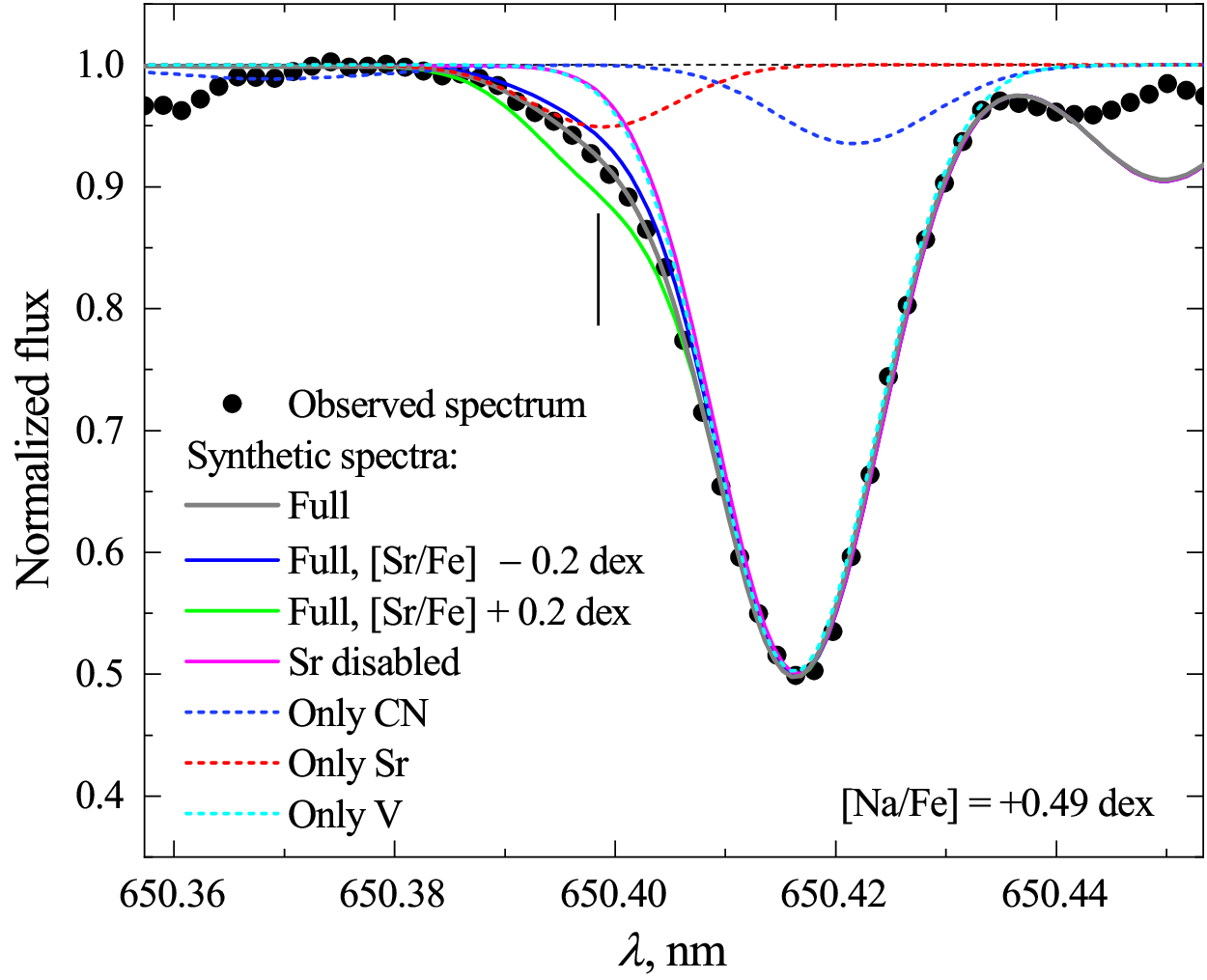}
	\caption{\ion{Sr}{i} line in the \UVES\ spectrum of the program star 27678 (dots) overlaid with the synthetic spectrum (solid gray line) that was calculated using the \ATLAS\ model with $\Teff=3958$\,K, $\log g=0.87$, $\FeH=-0.79$, $\xi_{\rm micro}=1.56$\,km\,s$^{-1}$, $\xi_{\rm macro}=6.20$\,km\,s$^{-1}$, [Sr/Fe] = 0.02\,dex. The other elemental abundances are scaled to solar \citep{GS98}. The other lines show the synthetic spectra of the full synthetic spectrum with the strontium abundance lowered and enhanced by 0.2\,dex (solid blue and green lines, respectively), the full synthetic spectrum without the \ion{Sr}{i} 650.3991\,nm line (solid magenta line), the CN lines alone (dashed blue line), the \ion{Sr}{i} line alone (dashed red line), and the \ion{V}{i} 650.4165\,nm line alone (dashed cyan line).	
	}
	\label{fig:Sr-blends}
\end{figure}

The 1D~LTE Sr abundances were determined using a spectral synthesis method, with the \SYNTHE\ package \citep{S05} used to compute the synthetic line profiles. A single 650.3991\,nm \ion{Sr}{i} was used in the analysis because it was the only Sr line available for the abundance analysis in the \UVES\ spectra. This line is located on the wing of another much stronger \ion{V}{i} (650.4165\,nm) line. Therefore, we tried our best to evaluate the influence of various factors on the determined Sr abundance, especially the influence of the \ion{V}{i} and CN line blends (Appendix~\ref{app-sect:abund-CN-impact}).

\begin{figure}[tb]
	\centering
	\includegraphics[width=\hsize]{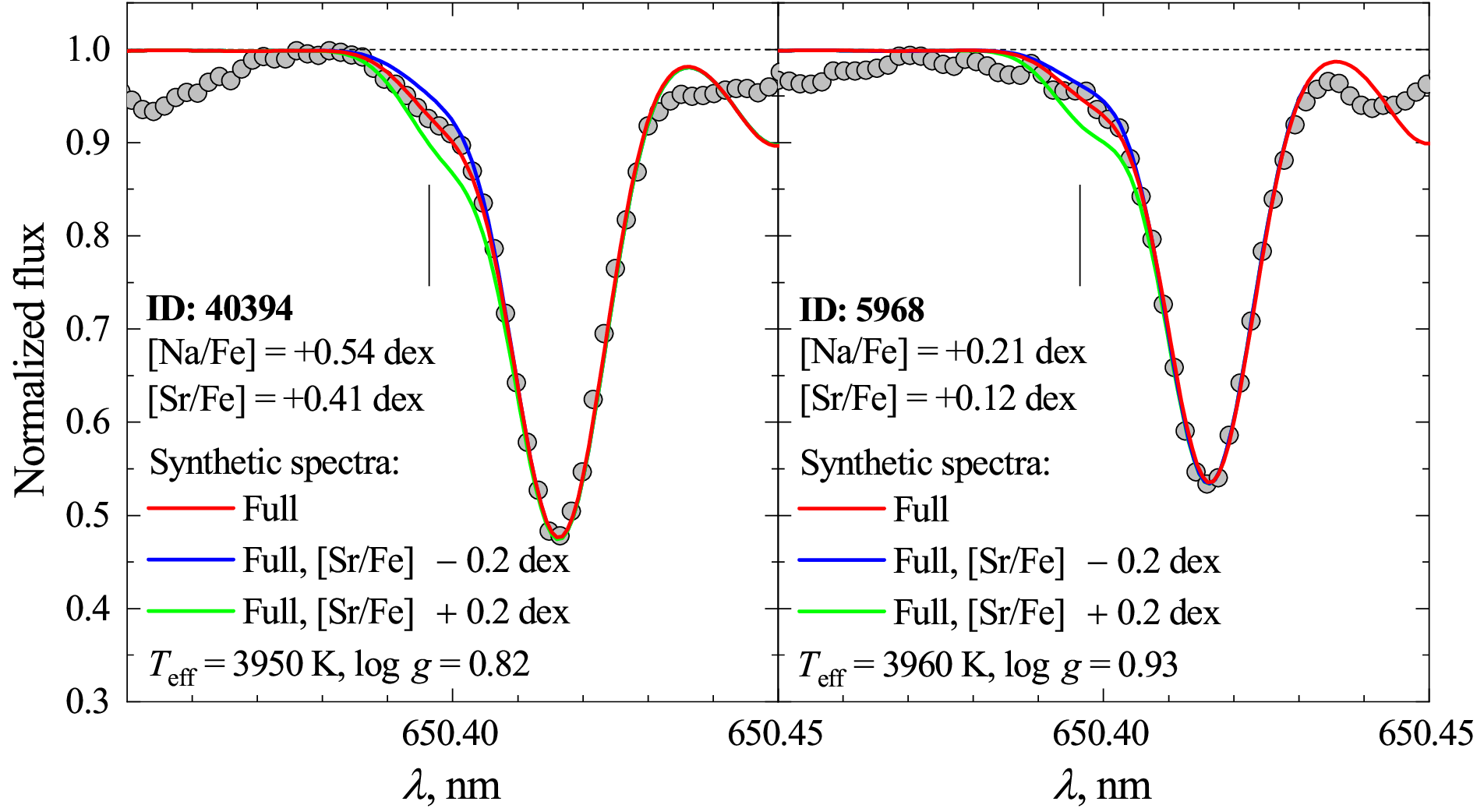}
	\caption{\ion{Sr}{i} line in the \UVES\ spectrum of two RGB stars (dots) in 47~Tuc: Na-rich (40394, 2P, left panel) and Na-poor (5968, 1P, right panel) overlaid with the full synthetic spectrum (solid red line). The identification numbers of each star, their atmospheric parameters, and the determined Sr and Na abundances are provided in each panel. The solid blue and green lines show the full synthetic spectrum, with the strontium abundance lowered and enhanced by 0.2\,dex, respectively.	
	}
	\label{fig:Sr-spectra-example}
\end{figure}

Because the analyzed \ion{Sr}{i} line is blended with the \ion{V}{i} (650.4165\,nm) line (the atomic line parameters for this line were taken from the VALD3 database and are shown in Table~\ref{tab:line-list}), the V abundance, together with the Sr abundance, was also determined iteratively in order to obtain a better synthetic line fit. An example of a best Sr line fit for target star 27678 is shown in Fig.~\ref{fig:Sr-blends}, together with the synthetic spectra computed with a $\pm 0.2$\,dex variation of the Sr abundance and different lines switched on or off.
Typical fits of the synthetic profiles to the \ion{Sr}{i} lines observed in the 1P and 2P target star spectra are shown in Fig.~\ref{fig:Sr-spectra-example}.

The mean V abundance in 47~Tuc determined by us, $\langle {\rm [V/Fe]} \rangle = 0.33\pm0.07$, differs from the those obtained by \citet{TSA14} ($\langle {\rm [Na/Fe]} \rangle = 0.17\pm0.09$) and \citet{Ernandes18} ($\langle {\rm [Na/Fe]} \rangle = -0.04\pm0.05$). Here, errors denote the standard deviation due to star-to-star scatter. Our target sample of 31 RGB stars contained 13 objects that were analyzed by \citet{TSA14}.
For this sample of 13 stars, we determined a mean V abundance $\langle {\rm [V/Fe]} \rangle = 0.31\pm0.07$, but the \Teff\ used by \citet{TSA14} was lower by 62\,K on average.
When we take into account the difference in \Teff, as well as the fact that the \ion{V}{i} lines that we used give systematically higher V abundance ($\sim 0.1$\,dex), our V abundance becomes lower by about 0.15\,dex (see Table~\ref{tab:Vanadium} in Appendix~\ref{app-sect:vanadium-tests}).
 
This brings it in line with the [V/Fe] values obtained by \citet{TSA14}. A similar case is seen with the [V/Fe] abundance ratios obtained by \citet{Ernandes18}, who used different log \textit{gf} values, Fe and V solar abundances, and Fe abundances of individual target stars. With these parameters from \citet{Ernandes18} our V abundances became lower by $\sim0.3$\,dex, which makes them comparable with those determined by \citet{Ernandes18} (see Appendix~\ref{app-sect:vanadium-tests} for more details).
	
The obtained Sr abundance shows no variation with the effective temperature and/or microturbulence velocity (Fig.~\ref{fig:Teff_vmic_COMBO}). The mean [Sr/Fe] ratio determined in 31 RGB stars is $\langle {\rm [Sr/Fe]} \rangle = 0.18\pm0.08$ (the error denotes the standard deviation due to the star-to-star abundance scatter).

\begin{figure}[tb]
	\centering
	\includegraphics[width=\hsize]{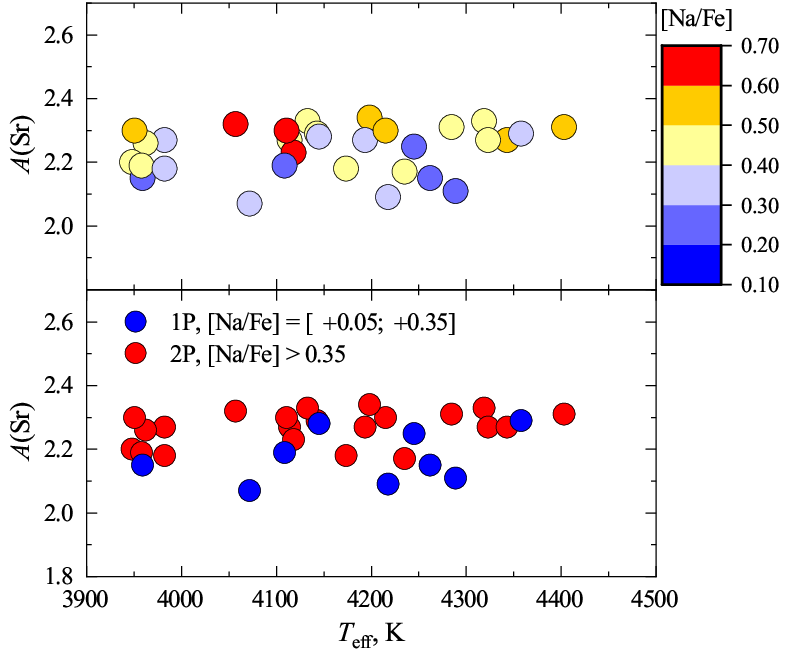}
	\caption{Strontium abundance in the target stars plotted vs. effective temperature. Top: Stars are marked in different colors according to their sodium abundances. Bottom: Blue circles show 1P stars (${\rm [Na/Fe]} = [{\rm [Na/Fe]}_{\rm min}, {\rm [Na/Fe]}_{\rm min}+0.3]$), and red circles show 2P stars (those with higher Na abundances). The minimum sodium value, [Na/Fe]$_\mathrm{min}=0.05$, was taken from \citet{KDK22}.}
	\label{fig:ASr_withNa_combo}
\end{figure}

In order to check for any pattern of population distribution in the effective temperature range, we marked the sodium abundances of individual stars in the \emph{A}(Sr) vs. \Teff\ plane as shown in the top panel of Fig.~\ref{fig:ASr_withNa_combo}.  Clearly, the first and second population stars are evenly distributed throughout the entire \Teff\ range. We further separated target stars into two populations, where, following \citet{CBG09a}, the 1P stars were defined as those with ${\rm [Na/Fe]} = [{\rm [Na/Fe]}_{\rm min}, {\rm [Na/Fe]}_{\rm min}+0.3]$, and the 2P stars with higher Na abundances.
The [Na/Fe]$_\mathrm{min}=0.05$ was determined using the Na NLTE abundances from the sample of 237 RGB stars analyzed in \citet{KDK22}.

Again, the stars belonging to the two populations are evenly distributed in the entire \Teff\ range (Fig.~\ref{fig:ASr_withNa_combo}, bottom panel).

The average strontium abundance measurement error, $\sigma$$_\mathrm{i}$, is $\sim$0.11\,dex (see Appendix~\ref{app-sect:abund-errors} for more detail). The strontium abundances determined for each individual star are provided in Table~\ref{app-tab:data-list}.

\subsection{Non-local thermodynamic equilibrium corrections for strontium}

To the best of our knowledge, no NLTE corrections for the \ion{Sr}{i} 650.3991\,nm are available from the literature, but there have been studies where corrections have been determined for other \ion{Sr}{i} and \ion{Sr}{ii} lines. In case of \ion{Sr}{i}, NLTE corrections are available only for the \ion{Sr}{i} 460.7331\,nm line and are about $\sim0.2$\,dex for stars with a metallicity and effective temperature similar to those of our target stars \citep{Bergemann2012}.
The dominating source of NLTE effects for the resonant \ion{Sr}{i} 460.7331\,nm line is overionization. However, because of the relatively low-excitation energy of the \ion{Sr}{i} 650.3991\,nm line  ($>2$\,eV), the NLTE effects are expected to be small.
In case of the \ion{Sr}{ii} lines, the NLTE corrections are usually negligible for stars with [Fe/H] > $-$2 \citep[see e.g.][]{Bergemann2012,Hansen2013}.
The NLTE corrections were calculated for Sr lines other than those used in this paper in several previous studies, however, the atmospheric parameters and metallicities of the stars studied there were very different from those of our target stars (e.g., \citealt{Andrievsky11}, \citealt{Zhao16}, \citealt{Mishenina19}, \citealt{Korotin20}, \citealt{Lardo21}, \citealt{Mashonkina22}). 

Importantly, because our target stars fall into a relatively narrow \Teff\ range ($\sim400$\,K) and are characterized by the same metallicity and very similar Sr abundances, nearly identical NLTE Sr abundance corrections for all target stars are also expected. Because of this, the Sr--Na correlation discussed in Sect.~\ref{sect:Sr-Na-correlation} should be unaffected by NLTE effects.

\subsection{The impact of uncertainty in the microturbulence velocity on the derived Sr abundance}

In order to determine whether the uncertainty of the microturbulence velocity has any impact on the derived Sr abundances, we performed two tests. In the first test, we calculated the synthetic profile of the Sr spectral line with the atmosphere parameters and Sr abundance values of star 2767 (unaltered profile). Then, two more synthetic Sr line profiles were calculated with the microturbulence velocity values adjusted by $\pm$\,0.2\,km~s$^{-1}$ (here we used the RMS variation of microturbulence velocity, $\xi_\mathrm{micro} = \pm$ 0.2\,km~s$^{-1}$, as a representative uncertainty in $\xi_\mathrm{micro}$). To match the latter two line profiles with the unaltered profile, the Sr abundance used to compute them would generally need to be adjusted. In reality, however, the adjustments needed were negligible ($<0.01$\,dex), and we therefore concluded that uncertainty in the microturbulence velocity had no impact on the determined Sr abundances.

The second test was the same as the first, except that the synthetic spectra were calculated with all spectral lines enabled, including those of the CN molecule. The results of this test agree with those of the first test: No changes were needed in the Sr abundances to match the unaltered profile with the synthetic spectra that were computed with the microturbulence velocity modified by $\pm$\,0.2\,km~s$^{-1}$.

\subsection{Uncertainty of the Sr abundances due to the effects of convection}

\begin{table}%
	\centering
	\caption{Atmospheric parameters and $\Delta_\mathrm{3D-1D~LTE}$ abundance corrections for the \ion{Sr}{I} line.}
	\label{tab:3D-1D} %
	\renewcommand{\arraystretch}{0.72}
	\scalebox{0.85}[0.85]{
		\begin{tabular}{ccccc}%
			\hline\hline\noalign{\vskip2.3pt} %
			\Teff,\,K & \logg & \ensuremath{\left[\mathrm{M}/\mathrm{H}\right]} & $\xi_{\rm micro}$,\,km s$^{-1}$ & $\Delta_\mathrm{3D-1D~LTE}$\\
			\hline\noalign{\vskip3pt}
			4040 & 1.5 & $-$1.0 & 1.5 & +0.05\\
			4485 & 2.0 & $-$1.0 & 2.0 & +0.01\\
			4492 & 2.5 & $-$1.0 & 2.0 & 0.00\\
			\hline\noalign{\vskip3pt}
		\end{tabular} }
	\end{table}
	
To date, the influence of convection on the formation of the \ion{Sr}{I} lines has not been explored. Due to the low-ionization potential of \ion{Sr}{I}, 5.695\,eV, neutral Sr is a minority species throughout a large part of our target star atmospheres. This could result in a significant sensitivity to convection because the horizontal temperature fluctuations associated with convection may notably impact the concentration of \ion{Sr}{I}, especially within the downdrafts located between the granules. This in turn may affect the \ion{Sr}{I} line strength and Sr abundances when the latter are determined using the 1D hydrostatic model atmospheres.

To assess the role of these effects, we computed the \ion{Sr}{i} 650.3991\,nm line profiles using the 3D hydrodynamical \COBOLD\ \citep{Freytag2012} and 1D hydrostatic \LHD\ model atmospheres \citep{CLS08}. The model atmospheres were computed for three sets of atmospheric parameters bracketing those of the target RGB stars (Table~\ref{tab:3D-1D}). Both the  3D and 1D model atmospheres were computed using identical chemical composition, opacities, and equations of state (see our similar previous studies, e.g., \citealt{DKS13} or \citealt{KDK16}, for more details regarding the method, model atmosphere, and spectral synthesis calculations). To compute synthetic line profiles, we used the spectral line parameters from Table~\ref{tab:line-list}, as well as the average microturbulence velocities and Sr abundances determined in our target stars with similar atmospheric parameters. Both the 3D and 1D line profiles were computed under the assumption of local thermodynamic equilibrium (LTE).

The obtained results indicate that the 3D--1D~LTE corrections are small ($<0.05$\,dex) and that their dependence on the atmospheric parameters (e.g., \Teff) is very weak. Furthermore, due to the weakness of the observed lines, the selection of the microturbulence velocity would not significantly affect the abundance corrections.

We stress that the obtained 3D--1D~LTE abundance corrections should not be directly applied to the measured Sr abundances. In the atmospheres of cool giants, both the 3D and NLTE effects may significantly affect the line formation. Thus, ideally, both the 3D and NLTE effects should be taken into account in the spectral line synthesis computations used in the abundance determinations. Nevertheless, the size of the obtained 3D--1D~LTE corrections lends us confidence that the effects of convection on the Sr abundances determined in our target stars are insignificant.

\section{Results and discussion\label{sect:discussion}}

\subsection{Mean abundance of Sr in the RGB stars of 47~Tuc}

\begin{table}[tb]
	\begin{center}
		\caption{Minimum, maximum, and mean Fe, Na, and Sr abundance ratios determined in the sample of 31 RGB stars in 47~Tuc.
			\label{tab:abundance-ranges}}
		\resizebox{\hsize}{!}{%
			\begin{tabular}{lcccccccc}
				\hline\hline
				\noalign{\smallskip}
				&  ${\rm [X/H]}_{\rm min}$  &  ${\rm [X/H]}_{\rm max}$  &  $\langle{\rm [X/H]}\rangle$  &  ${\rm [X/Fe]}_{\rm min}$  &  ${\rm [X/Fe]}_{\rm max}$  &  $\langle{\rm [X/Fe]}\rangle$ \\
				\hline\noalign{\smallskip}
				Fe~I (LTE)    &  $-0.98$  &  $-0.71$    &  $-0.81\pm 0.06$    &     --   &     --   &           --          \\				
				Na~I (NLTE)   &  $-0.60$  &  $-0.12$    &  $-0.29\pm 0.12$    &  $0.21$  &  $0.69$  &  $0.43\pm 0.12$  \\			
				Sr~I  (LTE)   &  $-0.78$  &  $-0.40$    &  $-0.63\pm 0.08$    &  $0.03$  &  $0.41$  &  $0.18\pm 0.08$\\			
				\hline
			\end{tabular}
		}
	\end{center}
	Note: Errors denote the standard deviation due to the star-to-star abundance variation.	
\end{table}

The only study of the Sr abundance in 47~Tuc performed so far was that of \citet{J04}, in which the authors determined Sr abundances in the atmospheres of eight subgiant branch (SGB) and three main sequence turn-off (TO) stars: $\langle {\rm [Sr/Fe]} \rangle_\mathrm{SGB} = 0.36 \pm 0.16$, and $\langle {\rm [Sr/Fe]} \rangle_\mathrm{TO} = 0.28 \pm 0.14$, respectively. The authors used two \ion{Sr}{ii} lines located at 407.7709 and 421.5519\,nm. The average abundances obtained in \citet{J04} are noticeably different from the average Sr abundance determined by us in 31 RGB stars, $\langle {\rm [Sr/Fe]} \rangle = 0.18 \pm 0.08$\,dex (Table~\ref{tab:abundance-ranges}; the error denotes the standard deviation due to the star-to-star abundance scatter). This disparity could arise from a range of factors, including those associated with the abundance analysis (e.g., employing a larger sample size in our study, variations in the placement of the continuum level, or use of different log\textit{gf} values in \citealt{J04}). Additionally, real discrepancies in the Sr abundance among the TO/SGB and RGB stars could also contribute to this abundance difference.

\begin{figure}[tb]
	\centering
	\includegraphics[width=\hsize]{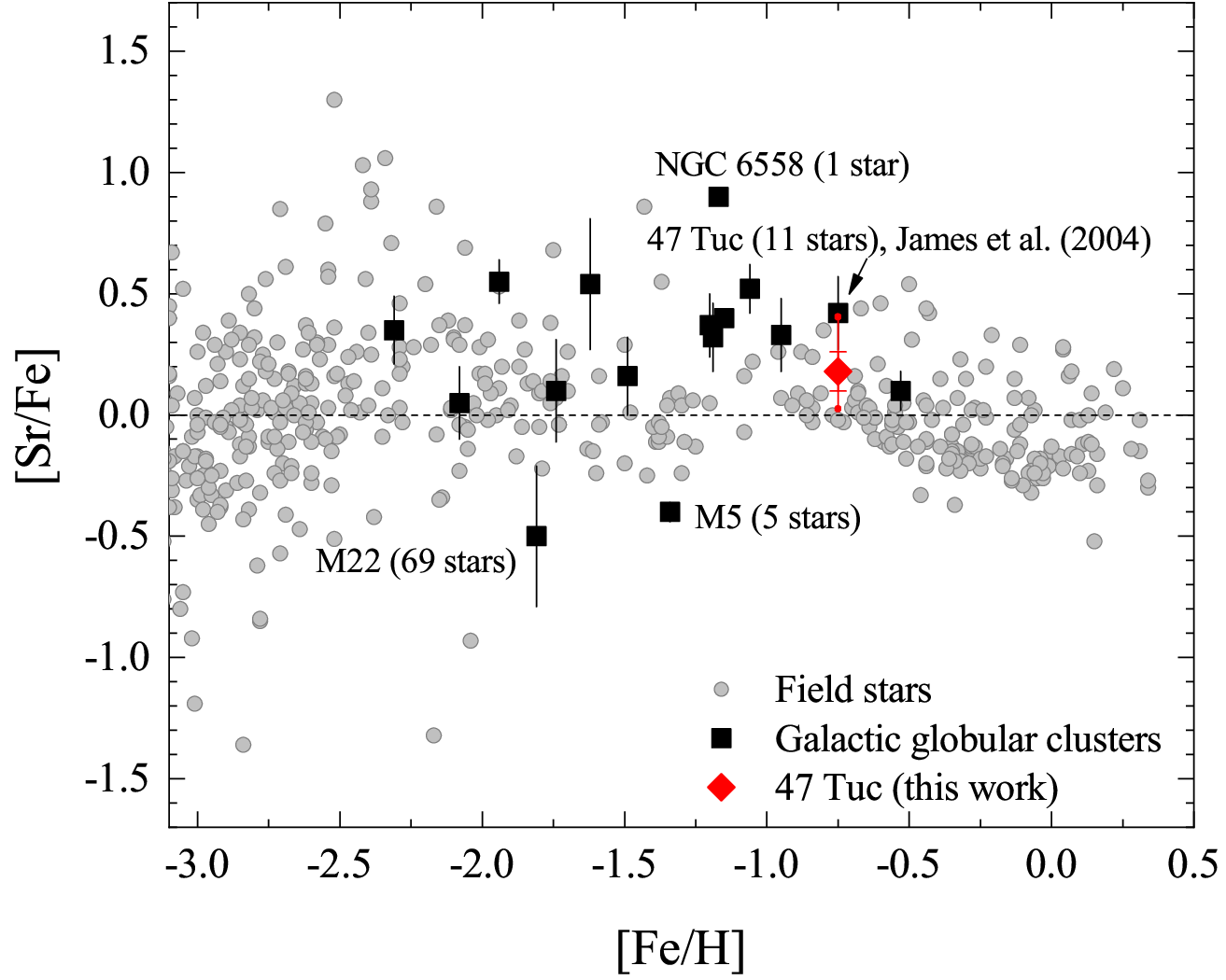}
	\caption{ Strontium-to-iron ratios in the Galactic globular clusters and field stars. The observational data for field stars are taken from \citet{Lai2008}, \citet{Ishigaki2013}, \citet{Roederer2014}, \citet{BB16}. The vertical black lines and the vertical red line with the horizontal line at the end show the standard deviation of the mean due to star-to-star abundance scatter. The vertical red line with dots at the end shows the minimum and maximum values. }
	\label{app-fig:GCSR}
\end{figure}

The mean Sr abundance in 47~Tuc determined in this work is well within the range of the Sr abundance variation observed in the Galactic field stars at the metallicity of 47~Tuc (Fig.~\ref{app-fig:GCSR}). 

\subsection{Possible Sr--Na abundance correlation?}\label{sect:Sr-Na-correlation}

\begin{figure}[tb]
	\centering
	\includegraphics[width=\hsize]{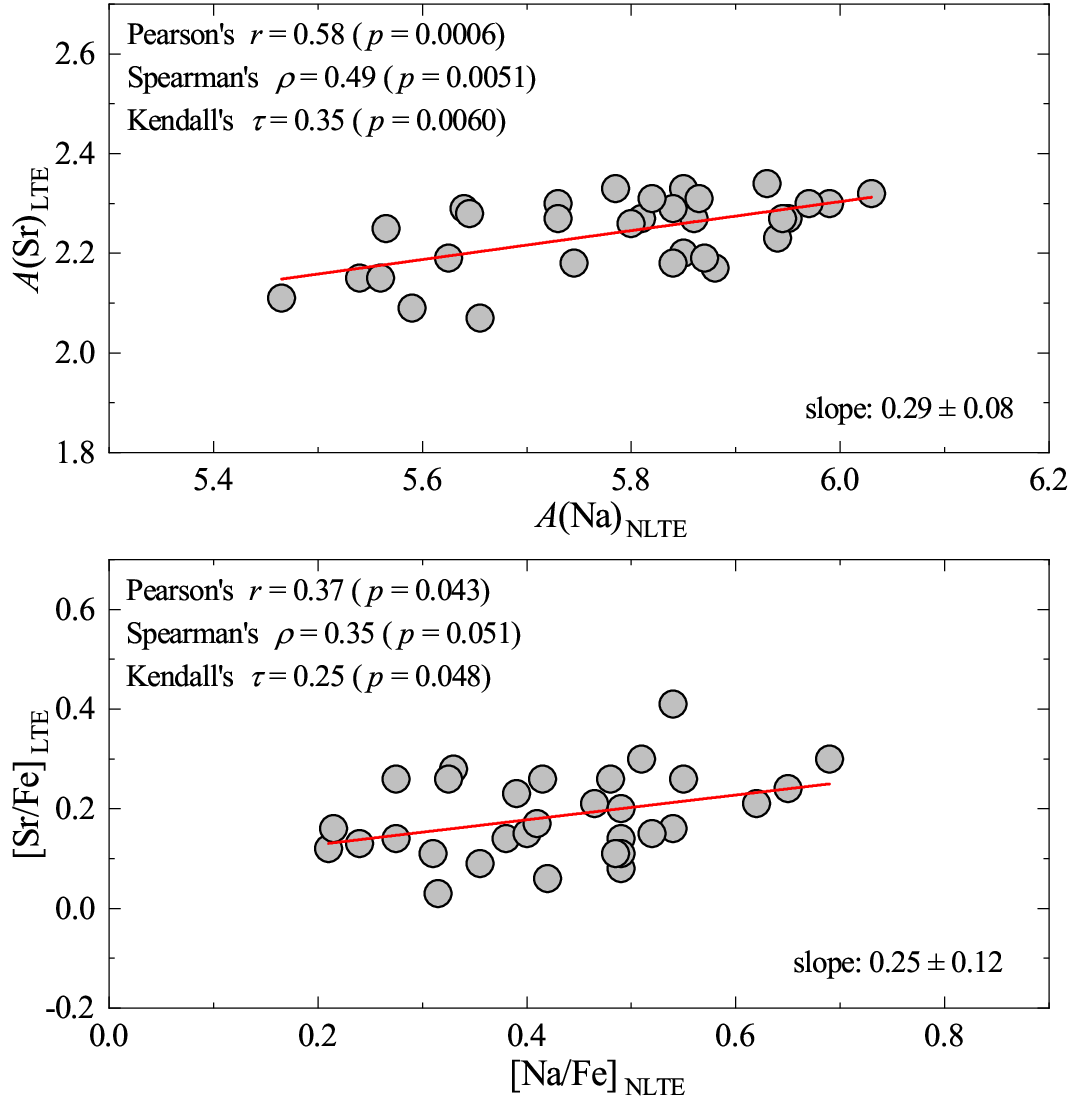}
	\caption{Sr abundance, \emph{A}(Sr), and [Sr/Fe] ratios determined in the target stars and plotted vs. the Na abundance, \emph{A}(Na) and [Na/Fe] ratio, respectively.}
	\label{fig:sprocess-vs-NaFe}
\end{figure}

Our data suggest weak correlations between (a) the Sr and Na abundances, and (b) the [Sr/Fe] and [Na/Fe] abundance ratios (Fig.~\ref{fig:sprocess-vs-NaFe}). Assuming the null-hypothesis that there is no Sr--Na correlation, the Pearson parametric as well as the Spearman and Kendall non-parametric tests show that the probability, $p$, to obtain the corresponding correlation coefficients, $r$, in our data set is never $p>0.05$ (see the $p$ values in Fig.~\ref{fig:sprocess-vs-NaFe}). This may suggest that the 2P stars in 47~Tuc are slightly enhanced in Sr. 

\begin{figure}[tb]
	\centering
	\includegraphics[width=\hsize]{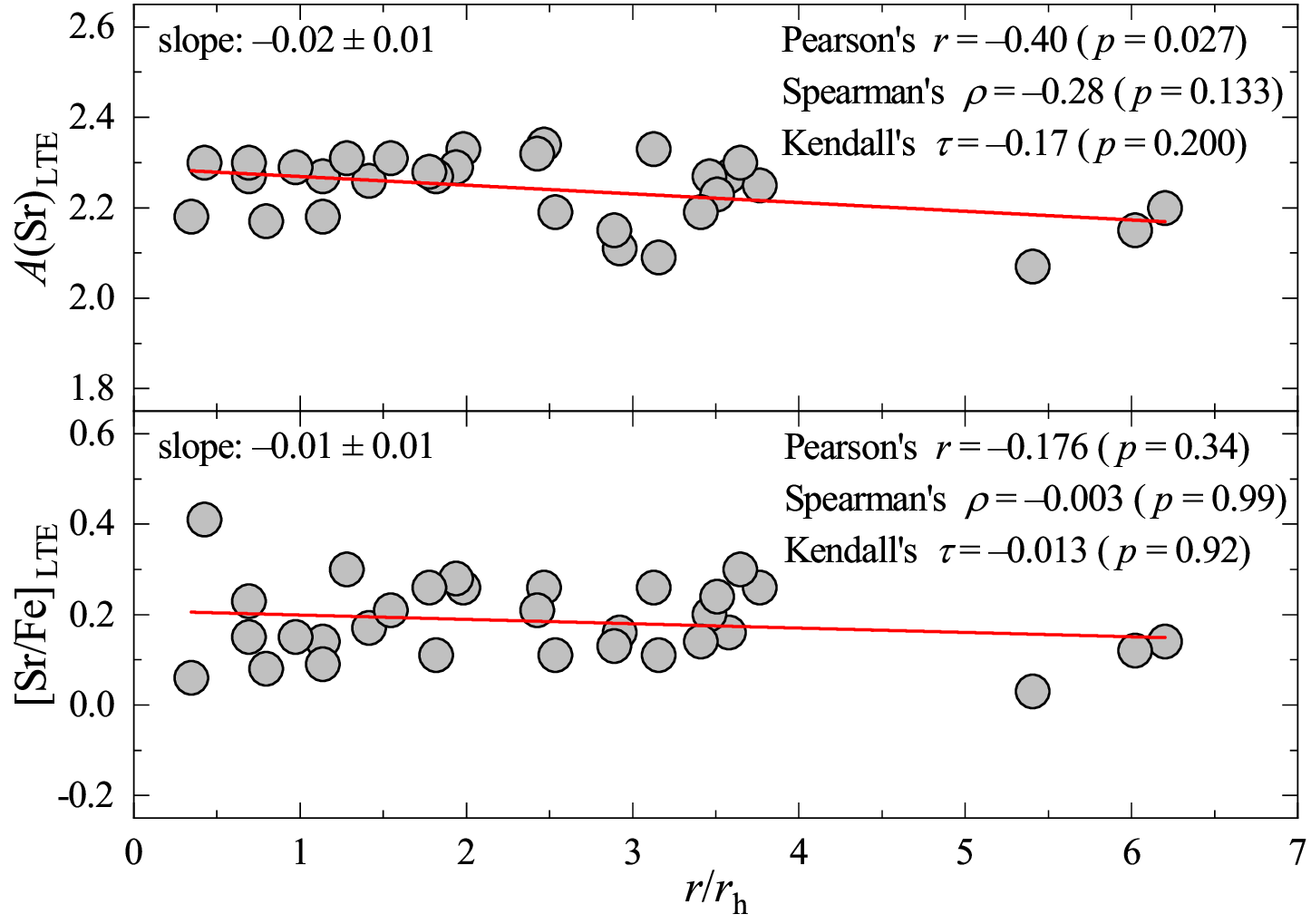}
	\caption{Abundance of Sr plotted vs. the relative distance from the cluster center. }
	\label{fig:SrFe-vs-rrh}
\end{figure}

\begin{figure}[tb]
	\centering
	\includegraphics[width=\hsize]{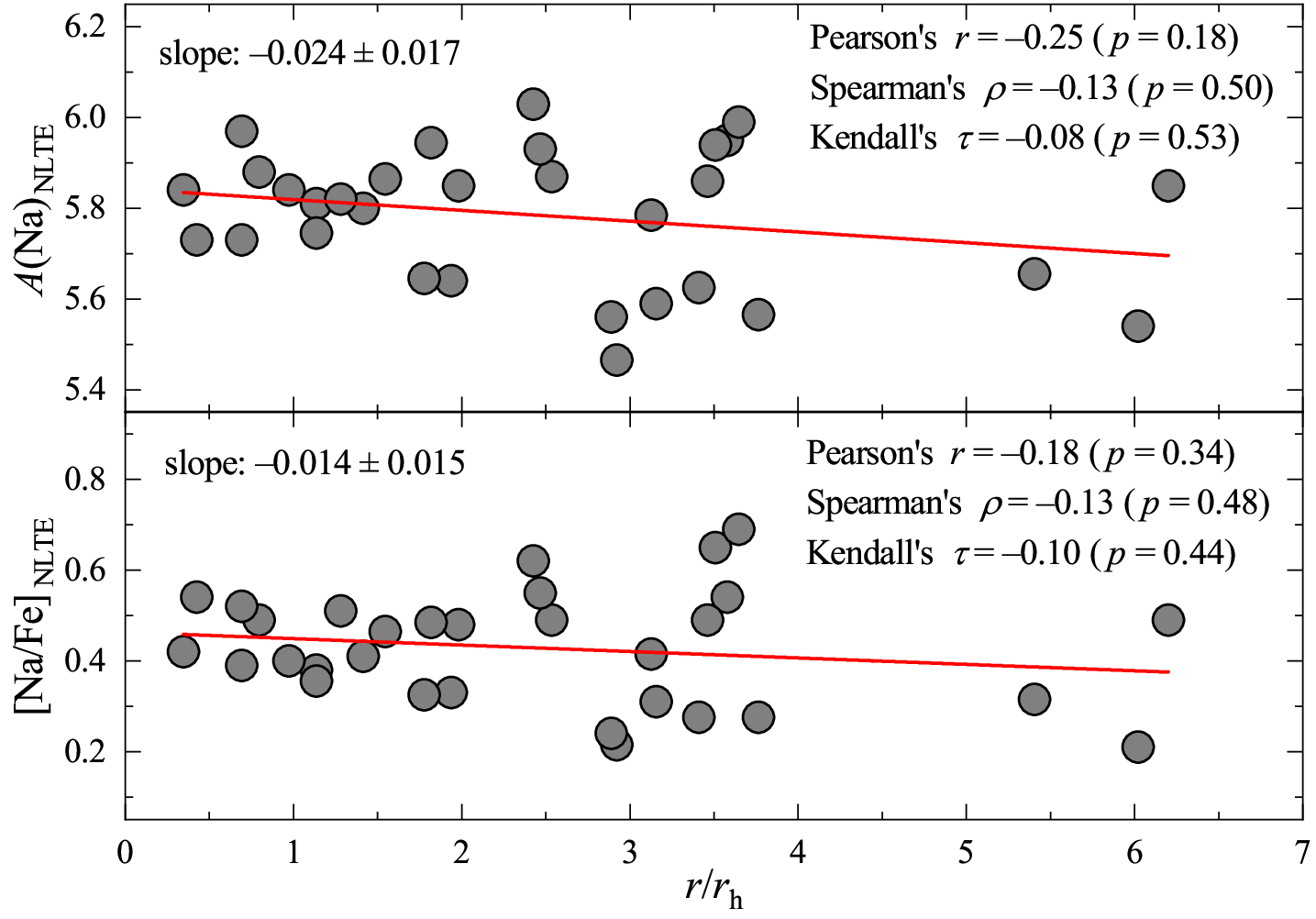}
	\caption{Abundance of Na plotted vs. the relative distance from the cluster center. }
	\label{fig:NaFe-vs-rrh}
\end{figure}

On the other hand, the statistical tests do not corroborate an Sr--$r/r_{\rm h}$ anticorrelation (see Fig.~\ref{fig:SrFe-vs-rrh} ;here, $r$ is the projected distance from the cluster center of a given target star, and $r_{\rm h}=174^{\prime\prime}$ is the half-light radius of 47~Tuc taken from \citealt{TDK93}). An anticorrelation like this would be expected if an Sr--Na correlation existed: In 47~Tuc, as well as in other GGCs, 2P stars (Na, Sr rich) tend to concentrate toward the cluster center. The Na--$r/r_{\rm h}$ anticorrelation in 47~Tuc has been confirmed in several studies (see, e.g., \citealt{KDK22} and references therein). However, no such anticorrelation is seen in our data (Fig.~\ref{fig:NaFe-vs-rrh}). Perhaps the most plausible explanation for this discrepancy is that the RGB star sample used in our study is simply too small to allow a reliable detection of the Sr--$r/r_{\rm h}$ anti-correlation. 

\begin{figure}[tb]
	\centering
	\includegraphics[width=\hsize]{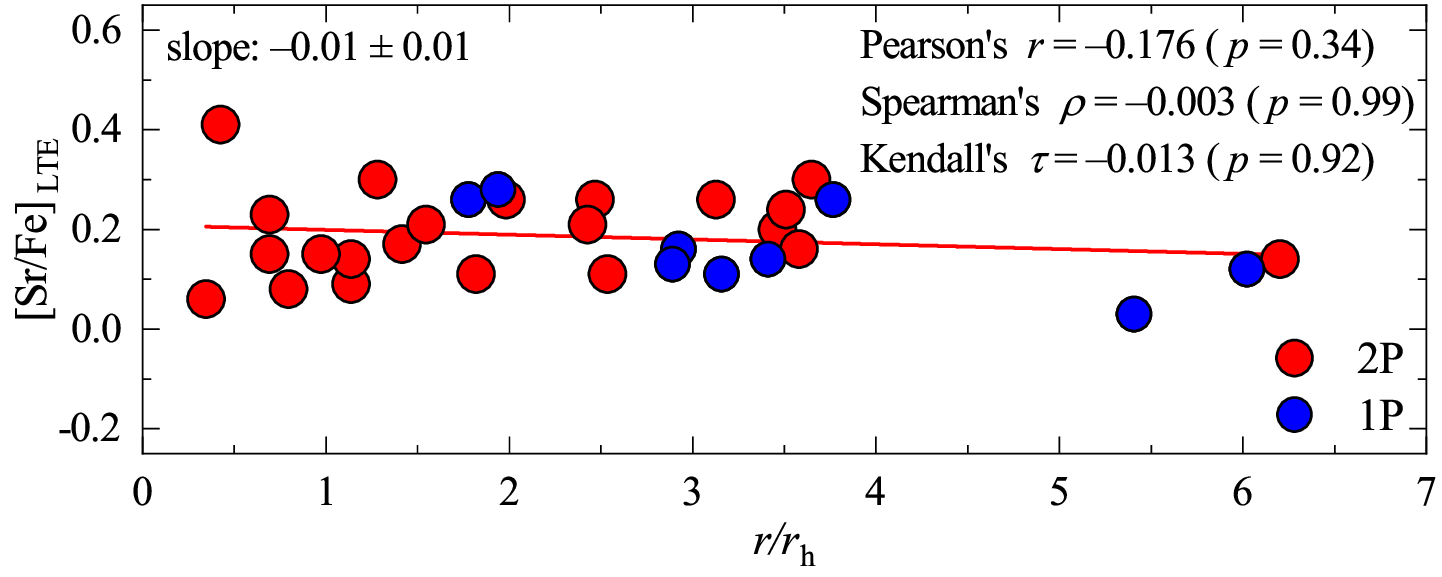}
	\caption{Abundance of Sr plotted vs. the relative distance from the cluster center. The stars are marked in different colors according to their population: blue shows first population stars (${\rm [Na/Fe]} = [{\rm [Na/Fe]}_{\rm min}, {\rm [Na/Fe]}_{\rm min}+0.3]$), and red shows second population stars (those with higher Na abundances). The minimum sodium value, [Na/Fe]$_\mathrm{min} = 0.05$, is taken from \citet{KDK22}. }
	\label{fig:NEW_COMBO_Sr_vs_rrh_1P2P}
\end{figure}

\begin{figure}[tb]
	\centering
	\includegraphics[width=\hsize]{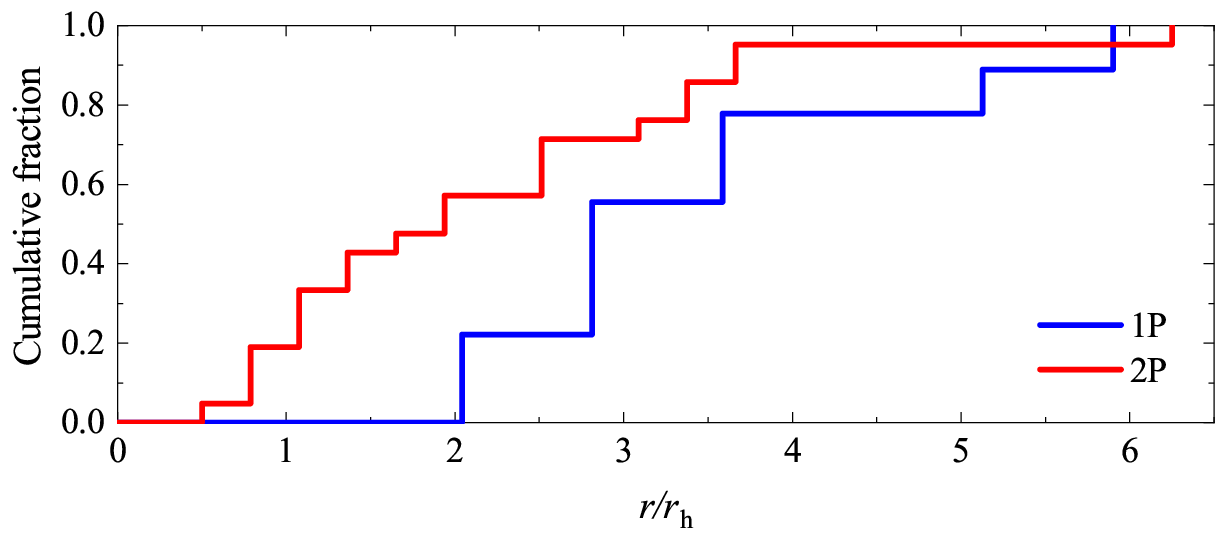}
	\caption{Cumulative number fraction of the stars in the the Na-rich (red line) and Na-poor (blue line) populations plotted vs. the relative distance from the cluster center. The results of the K–S test show that the probability of the 1P and 2P groups to be drawn from the same sample is $p$ = 0.052.}
	\label{fig:CumulativeFraction2}
\end{figure}

On the other hand, our data seem to suggest that the more Na-rich stars are in fact more concentrated toward the cluster center. This can by checked by using the  samples of Na-rich and Na-poor stars selected in Sect.~\ref{sect:abund-sr}. Both the visual inspection of their distribution versus the distance from the cluster center, $r/r_{\rm h}$ (Fig.~\ref{fig:NEW_COMBO_Sr_vs_rrh_1P2P}) and the results of the Kolmogorov-Smirnov (K-S) test of the cumulative distributions of the 1P and 2P stars versus $r/r_{\rm h}$ (Fig.~\ref{fig:CumulativeFraction2}) suggest that the Na-rich population is indeed concentrated toward the cluster center. In particular, the probability that the cumulative  distributions of the Na-rich and Na-poor stars are identical is $p\lesssim0.05$, as estimated from the K-S test, whereas no statistical evidence for this can be detected in the Na--$r/r_{\rm h}$ plane, most likely because the sample is too small for us to obtain a reliable estimate in this plane.

\subsection{Production of s-process elements in the 2P polluters?}\label{sect:s-process-polluters}

The Sr results obtained above suggest that the polluters that have enhanced (or diluted) the 2P stars with light elements may have produced a noticeable amount of s-process elements as well. A small but statistically significant 2P--1P difference in the abundance of another first s-process peak element, Zr, and a Zr--Na correlation was found in our earlier analysis of Zr in 47~Tuc \citep{KDK22}. The results of the two studies may therefore suggest that the 2P stars in 47~Tuc could have been enriched in the first-peak s-process elements. This assertion clearly has to be confirmed by the analysis of Sr abundance in a larger sample of stars, as well as by the analysis of other s-process elements both in this and other GGCs, especially Y and Rb, which belong to the same (first) s-process peak as Sr and Zr.

The analysis of the heavier s-process elements performed by different authors has so far yielded more mixed results. In their study of a relatively small sample of 13 RGB stars in 47~Tuc, \citet{TSA14} detected no Ba or La abundance correlations with Na. No La--Na correlation has been observed in 123 RGB stars of this cluster by \citet{CPJ14}. In our recent analysis of the Ba abundance in 261 RGB stars of 47~Tuc, we found no correlation between the Ba and Na abundances \citep{DKK21}. While Ba and La belong to the second s-process peak, the theoretical abundance yields from AGB and FRMS stars suggest that Ba and La should have been synthesized together with the first s-process peak elements, for instance, Sr, Zr (see, e.g., \citealt{CBG09a, LC18}). On the other hand, in their recent studies of 47~Tuc and several other GGCs, \citet{FBB2021,FVG2022} suggested the existence of Ce--N and Ce--Al correlations. The absence of a 2P--1P difference for Ba and La is difficult to reconcile with a possible Ce--N correlation because all these elements belong to the the same second s-process peak and are expected to follow the same abundance patterns.

While the investigation of possible correlations between the abundances of Sr and those of C+N+O or Sr and Al would be interesting, only eight stars in our sample have C+N+O abundances determined by \citet{MMGH20}, while in the case of Al, this is true for only five stars. The number of stars is therefore too small to draw any reliable conclusions regarding any possible correlations.

\begin{figure}[tb]
	\centering
	\includegraphics[width=\hsize]{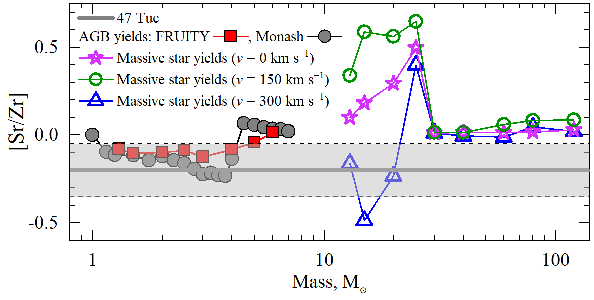}
	\caption{[Sr/Zr] ratio vs. stellar mass. Theoretical AGB yields: FRUITY \citep{CSP15}, and Monash \citep{Karakas18}. The massive star yields are taken from \citet{LC18}. The horizontal solid gray line shows the [Sr/Zr] abundance ratio of 47~Tuc, where the abundances of Sr were determined in this work and the abundances of Zr were taken from \citet{KDK22}. The shaded gray area shows the standard deviation due to the star-to-star Sr and Zr abundance scatter.   }
	\label{app-fig:iron-teff-vmic}
\end{figure}

A comparison of the mean Sr and Zr abundance ratio in 47~Tuc (the latter taken from \citealt{KDK22}) with the theoretical predictions of s-process nucleosynthesis in AGB and massive stars suggests that both elements could have been produced either in the AGB stars or (less conclusively) in high-mass ($M=10-20$\,M$_\odot$) rapidly rotating ($v_{\rm rot}=200-300$\,km\,s$^{-1}$) stars (Fig.~\ref{app-fig:iron-teff-vmic}). Clearly, more data on the s-process element abundances are needed to better constrain the possible polluters in the GGCs.

\section{Conclusions and outlook}\label{conclus}

The mean Sr abundance determined by us in 31 stars in 47~Tuc is $\langle {\rm [Sr/Fe]} \rangle = 0.18\pm0.08$ (Table~\ref{tab:abundance-ranges}; the error denotes the standard deviation due to star-to-star abundance scatter). This value is lower than the one obtained in the earlier study of \citet{J04}, who used significantly smaller samples of SGB stars (eight objects) and TO stars (three objects), $\langle {\rm [Sr/Fe]} \rangle_\mathrm{SGB} = 0.36 \pm 0.16$ and $\langle {\rm [Sr/Fe]} \rangle_\mathrm{TO} = 0.28 \pm 0.14$, respectively. The difference may result from a variety of factors, and it may be due to the significantly larger sample size in our study, the adoption of different atomic parameters (e.g., the $\log gf$ values), and so on (see Sect.~\ref{sect:abund-sr} for details).

Our analysis of the Sr abundance in 31 RGB stars in 47~Tuc suggests a weak Sr--Na abundance correlation. Together with a 2P--1P difference in Zr abundance and Zr--Na correlation observed in our earlier study \citep{KDK22}, this may suggest that the polluters that enhanced (or depleted) the 2P stars with light elements may have synthesized noticeable amounts of Sr and Zr as well. It is clearly very desirable to confirm the existence of the Sr--Na correlation using a larger sample of stars and other spectral lines of Sr. Because no 2P--1P abundance differences were detected in the earlier studies of the first-peak s-process elements in the GGCs, it would also be very interesting to verify whether these 2P--1P s-process differences exist for other first-peak s-process elements in this and other GGCs.

At the same time, the results for heavier s-process elements, such as Ba, La, and Ce, are more ambiguous, with possible 2P--1P differences observed for some elements (e.g., Ce, \citealt{FBB2021,FVG2022}), but not for others (e.g., Ba and La). It is important to stress that to the best of our knowledge, no studies of the third-peak s-process elements have been performed so far. These investigations are therefore extremely desirable as well.

Clearly, a further analysis of s-process elements in conjunction with theoretical predictions for different polluters is required to gain more information about the abundance patterns of s-process elements in different clusters and on the nature of the possible polluters. Even if the 2P--1P differences may be small in nature and the correlations between the s-process and light element abundances are not detected reliably, a confirmation or rejection of the abundance differences in larger samples of 2P and 1P stars would provide valuable additional information about the nature of the possible polluters and nucleosynthesis in the GGCs in general.

\begin{acknowledgements}
%
{We are very grateful to Tamara Mishenina and Sergey Andrievsky for useful comments during the preparation of the manuscript. Our study has benefited from the activities of the "ChETEC" COST Action (CA16117), supported by COST (European Cooperation in Science and Technology) and from the European Union’s Horizon 2020 research and innovation programme under grant agreement No 101008324 (ChETEC-INFRA). This work has made use of the VALD database, operated at Uppsala University, the Institute of Astronomy RAS in Moscow, and the University of Vienna. J.K. acknowledges support from European Social Fund (project No 09.3.3-LMT-K-712-19-0172) under grant agreement with the Research Council of Lithuania.}
\end{acknowledgements}




\begin{appendix}

\section{Archival spectra\label{app-sect:arch_spectra}}

To study the Sr abundance in 47~Tuc, we searched the ESO Advanced Data Product (ADP) archive for spectra that contain at least one Sr line that can be measured reliably. We found four programs (Table~\ref{tab:obs-journal}) that were performed using the VLT UT2 telescope and contained the \ion{Sr}{i} 650.3991\,nm line, which was subsequently used for the Sr abundance determination in our study.

The other Sr lines that were used to determine the strontium abundances in previous studies were either too weak or too noisy to allow a reliable abundance determination. These lines were the \ion{Sr}{i} 496.2259\,nm \citep{Monty23}, the \ion{Sr}{i} 650.3991\,nm \citep{Barbuy2014}, the \ion{Sr}{i}  655.0244\,nm \citep{Barbuy2018}, and the \ion{Sr}{i} 679.1016\,nm line \citep{Barbuy2014}. The  other candidate Sr lines that we tested were the \ion{Sr}{i} 487.2488\,nm, the \ion{Sr}{i}  640.8459\,nm, the \ion{Sr}{i} 654.6784\,nm, the \ion{Sr}{i} 661.7263\,nm, and the \ion{Sr}{i} 679.1016\,nm line. Unfortunately, they were all too weak to be observed in the \UVES\ spectra. 

Our search of the \UVES\ or \GIRAFFE\ observations in the ESO ADP archive yielded no spectra that would contain other Sr lines, such as the \ion{Sr}{i} 460.7331\,nm \citep{Bergemann2012}, the \ion{Sr}{i} 707.0070\,nm \citep{Yong2017}, the \ion{Sr}{ii} 416.1791\,nm, and the \ion{Sr}{ii} 430.5443\,nm line \citep{Andrievsky11}.

The spectra of the program in the \GIRAFFE\ archive, 089.D-0579(A) cover the \ion{Sr}{ii} 421.5519\,nm line, which was used earlier by \citet{Andrievsky11}. Unfortunately, the noise level in these spectra is too high and their spectral resolution is too low for a reliable abundance determination.

Given the lack of other observational material, we therefore relied on the \UVES\ spectra from the programs listed in Table~\ref{tab:obs-journal} to determine the Sr abundance using a single \ion{Sr}{i} 650.3991\,nm line.

\section{Cluster membership\label{app-sect:clust_memb}}

\begin{figure}[h!]
\centering
\includegraphics[width=\hsize]{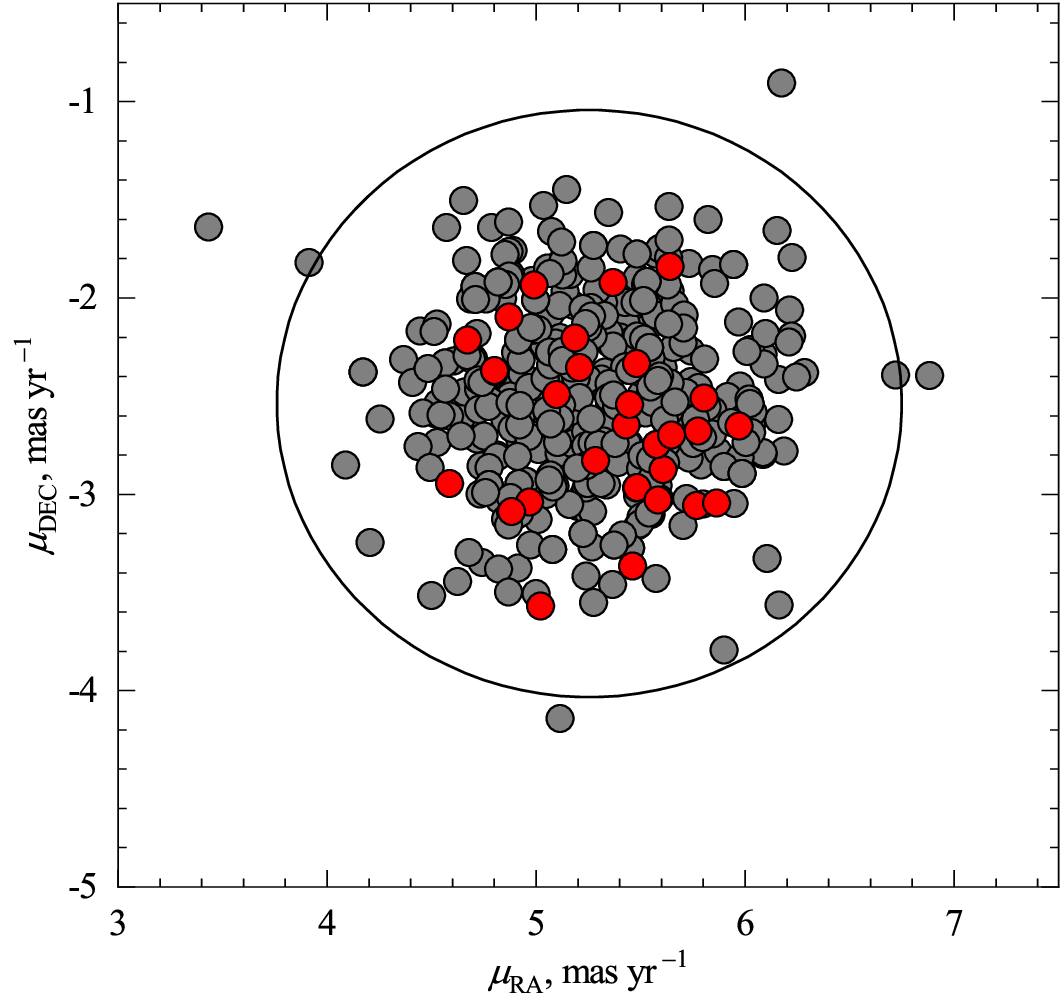}
\caption{Proper motions of the target RGB stars in 47~Tuc from the \textit{Gaia} EDR3 catalog \citep{GC23}. The red and gray symbols denote the stars analyzed in this work and those studied in \citet{KDK22}, respectively. The black circle encompasses the region with a radius of 1.5\,mas\,yr$^{-1}$ from the mean cluster proper motion.  }
\label{app-fig:PM}
\end{figure}

In order to verify the cluster membership of the target RGB stars, we compared their proper motions (taken from \textit{Gaia} DR3 catalog, \citealt{GC23}) with the mean proper motion of 47~Tuc. Following \citet{MM18}, the target star was assigned a cluster membership when its proper motion did not deviate by more than 1.5\,mas\,yr$^{-1}$ from the mean proper motion of 47~Tuc, $\mu_{\rm RA} = 5.25$\,mas\,yr$^{-1}$ and $\mu_{\rm DEC} = -2.53$\,mas\,yr$^{-1}$ \citep{BH19}. The result of this test showed that all selected target stars fulfilled this criterion (Fig.~\ref{app-fig:PM}).

\section{Errors in the derived abundances\label{app-sect:abund-errors}}

\begin{table*}[h!]
	\begin{center}
		\caption{Typical errors in the measured Fe, Na, and Sr abundances.
			\label{app-tab:errors_full}}
			\resizebox{12.0cm}{!}{%
		\begin{tabular}{cccccccccc}
			\hline\hline
			\noalign{\smallskip}
			Element&\textit{T}$_{\mathrm{eff}}$ & Star ID & $\sigma$(\textit{T}$_{\mathrm{eff}}$)  & $\sigma$(log\textit{g})  & $\sigma$($\xi_\mathrm{micro}$)  & $\sigma$(cont)  & $\sigma$(fit)  &$\sigma$$_\mathrm{i}$(total)  \\
			&	K &  &    dex     & dex                            &  dex  & dex & dex  & dex            \\
			\hline\noalign{\smallskip}
			\ion{Fe}{I} &	4057	&	s17657     	& 0.03	 	&	0.02 	&	0.14 	&  0.09 	& 0.02    	&	0.17   	\\
			&	4358	&	20885   	& 0.03	 	&	0.02 	&	0.12 	&  0.09 	& 0.03    	&	0.16   \\							
			\ion{Na}{I} &	4057	&	s17657     	& 0.06	 	&	0.01 	&	0.06 	&  0.03 	& 0.03    	&	0.10   	\\
			&	4358	&	20885   	& 0.06	 	&	0.01 	&	0.04 	&  0.03 	& 0.04    	&	0.09   	\\				                    		
			\ion{Sr}{I} &	4057	&	s17657     	& 0.06	 	&	0.00 	&	0.00 	&  0.01 	& 0.09    	&	0.11   	\\
			&	4358	&	20885   	& 0.05	 	&	0.00 	&	0.00 	&  0.02 	& 0.10    	&	0.11   	\\						            				
			\hline
		\end{tabular}
	}
	\end{center}
\end{table*}

The error in the abundance estimate due to the uncertainty in the effective temperature determination was obtained by computing new effective temperature values that were increased or decreased by an amount that corresponded to the uncertainty in the determination of the $V - I$ color index. For the latter, a conservative estimate of $\sigma(V) = \sigma(I) = 0.03$\,mag was used, which translates into an uncertainty of $\pm$\,60\,K in the effective temperature. The corresponding uncertainties in the abundance estimates are provided in Table~\ref{app-tab:errors_full}.

The error in the abundance arising due to the uncertainty in the determination of the surface gravity, \logg, was estimated by taking into account the uncertainties in the determination of  the effective temperature, luminosity, and stellar mass errors. The resulting error in \logg\ was $\pm0.2$ and its impact on the abundance error was negligible (Table~\ref{app-tab:errors_full}.). 

To estimate the error due to the uncertainty in the microturbulence velocity ($\xi_\mathrm{micro}$) determination, we used the RMS variation of microturbulence velocity as a representative uncertainty in $\xi_\mathrm{micro}$, namely, $\pm$\,0.2\,km\,s$^{-1}$.

The error in the continuum determination was estimated by computing the flux dispersion $\sigma_{cont}$ in the continuum as

\begin{equation}
err(cont) = \frac{\sigma_{cont}}{\sqrt{N}},
\end{equation} 

\noindent where $N$ is total number of wavelength points (usually $30 < N < 50$). The obtained value was then used to increase and decrease the continuum level, after which, the line profile fitting was repeated and the resulting difference in the abundance was further used as the abundance error due to the uncertainty in the continuum determination.

To estimate the errors arising in the process of the line profile fitting, we computed the RMS deviation between the observed spectral line profiles and the corresponding best-fit Gaussian profile. The obtained values were further used to increase or decrease the strength of the line profile to obtain the error in the determined abundances (Table~\ref{app-tab:errors_full}).

\section{CN impact on the derived abundances\label{app-sect:abund-CN-impact}}

The 650.3991\,nm \ion{Sr}{i} line used in our study is weakly blended by a nearby CN line (Fig.~\ref{fig:Sr-blends}). As shown by \citet{MMGH20}, the abundances of the CNO elements can differ strongly in the 1P and 2P stars in 47~Tuc. This may lead to overestimated Sr abundances in the 2P stars when this particular \ion{Sr}{i} line is used in the abundance determination. In turn, this may lead to a spurious correlations or anticorrelations between the abundance of Sr and those of the light chemical elements.

To estimate the influence of the CN blends on the determined Sr abundances, we performed the following test. Two second-population stars (IDs: 27678, s23821) with different $\Teff$ (4000\,K and 4400\,K) but identical [Na/Fe] abundances were chosen. We then derived the Sr abundances in the two stars using two different sets of CNO corresponding to the abundances of the 1P (CN-weak) and 2P (CN-strong) stars (selected from the distribution of the CNO abundances in a set of 82 stars in 47~Tuc from \citealt{MMGH20}), 

\begin{itemize}
	\item[$\bullet$]
	${\rm [C/Fe]}=+0.20$, ${\rm [N/Fe]}=+0.20$, ${\rm [O/Fe]}=+0.80$ (case A, CN-weak, 1P);
	\item[$\bullet$]
	${\rm [C/Fe]}=-0.80$, ${\rm [N/Fe]}=+1.70$, ${\rm [O/Fe]}=+0.10$ (case B, CN-strong, 2P).
\end{itemize}

We thus obtained two Sr abundance estimates in each star and computed the difference between the Sr abundances obtained using the case A and case B CNO abundances. The obtained results showed that the difference between the Sr abundances determined using the case A and case B CNO values were negligible and never exceeded 0.01\,dex. 

In addition, we also performed a second CN test because the Sr line is located on the wing of another much stronger V line.
In this second test, we calculated synthetic spectra with only the Sr and CN lines enabled, using the same stellar and abundance parameters as described above. These synthetic spectra were then used as real spectra from which we determined the equivalent widths of the Sr lines (i.e., Sr line + CN blend). The determined equivalent widths were then further used to obtain the Sr abundances with \WIDTH\ package. The results of this test showed that the difference between Sr abundances when CNO abundances are set with the case B and case A values are not larger than 0.02\,dex. 

Taking into account the results obtained in the two tests, we therefore conclude that the influence of CN blends has a negligible impact on the determined Sr abundances.

\section{Vanadium abundance\label{app-sect:vanadium-tests}}

As we mentioned in Sect.~\ref{sect:abund-sr}, together with the Sr abundances, we also determined the V abundances from a single 650.4165~nm \ion{V}{i} line. The average V abundance ratio derived in this work from 31 RGB stars using this line is $\langle {\rm [V/Fe]} \rangle = 0.33 \pm 0.07$ (the error denotes the standard deviation due to the star-to-star abundance variation). This value is higher than the value obtained by \citet{Ernandes18} for 5 RGB stars in this same cluster ($\langle {\rm [V/Fe]} \rangle = -0.04 \pm 0.05$), but in that work, eight other \ion{V}{i} lines were used, which were different from the line used in this work. Our spectra include these eight lines, and therefore, we derived the V abundances from these eight lines in 3 RGB stars from our star sample in order to determine whether this large difference in the determined V abundances remained. The results of this test show that even when using the same eight lines that were used in \citet{Ernandes18}, we obtain [V/Fe] abundance ratios of about 0.30\,dex higher than \citet{Ernandes18} (columns 2, 4, and 6 in Table.~\ref{tab:Vanadium}). Interestingly, these eight lines produce up to 0.10\,dex lower V abundance than the original 650.4165~nm \ion{V}{i} line, which was used for 31 target stars in this work. The most likely cause is a blend with CN line (see Fig.~\ref{fig:Sr-blends}).

Furthermore, in order to determine the [V/Fe] abundance ratios, \citet{Ernandes18} used different log\textit{gf} values, Fe and V solar abundances (higher by 0.07 and lower by 0.05\,dex respectively) and Fe abundances of individual target stars (higher by 0.13\,dex on average); when we recalculated the abundances with the parameters used in their work (also keeping in mind that the 650.4165~nm \ion{V}{i} line produces an abundance that is higher by up to 0.10\,dex than the other eight \ion{V}{i} lines, as shown in Table.~\ref{tab:Vanadium}), our determined [V/Fe] abundance ratios became lower by up to 0.30\,dex (columns 3, 5, and 7 in Fig~\ref{tab:Vanadium}). This in turn brings the average [V/Fe] abundance ratio very close to the one determined in \citet{Ernandes18}. However, we do not recommend using the V abundances obtained in this work from the 650.4165\,nm \ion{V}{i} spectral line for any analysis because this line is blended with the CN line, which leads to an incorrect V abundance assessment. The V abundances determined in this work were a byproduct of the Sr abundance analysis, as the analyzed \ion{Sr}{i} 650.3991\,nm line is blended with the \ion{V}{i} 650.4165\,nm line and we had to determine the V abundance using the \ion{V}{i} 650.4165\,nm line in order to fit the total synthetic spectrum to the observed spectrum.

\begin{table*}[tb]
	\begin{center}
		\caption{[V/Fe] abundance ratios determined in the sample of three RGB stars from our target sample with the highest [Na/Fe] abundance ratios (columns 2, 4, and 6). The star IDs are given in the first row of the table. The first column shows the central wavelengths of the \ion{V}{i} spectral lines. Columns 3, 5, and 7 (marked with an asterisk) show the [V/Fe] abundance ratios that were recalculated with log \textit{gf} values, Fe and V solar abundances, and the Fe abundances of individual target stars taken from \citet{Ernandes18}.
			\label{tab:Vanadium}}
		\resizebox{13.0cm}{!}{%
			\begin{tabular}{l|cr|cr|cr}
				\hline\hline
				\noalign{\vskip 1mm}
				& \multicolumn{2}{c}{00241344-7211263}   &     \multicolumn{2}{c}{s17657}   &  \multicolumn{2}{c}{s5270}      \\
				$\lambda$, nm  &  ${\rm [V/Fe]}$  &  ${\rm [V/Fe]}_{\rm *}$  & ${\rm [V/Fe]}$  &  ${\rm [V/Fe]}_{\rm *}$  &  ${\rm [V/Fe]}$  &  ${\rm [V/Fe]}_{\rm *}$ \\
				\noalign{\vskip 0.6mm}
				\hline
				\noalign{\vskip 1mm}
				570.3560    &  $0.21$  &  $-0.06$    &  $0.28$    &  $0.13$  &  $0.27$  &  $0.01$       \\				
				608.1440    &  $0.32$  &  $0.08$     &  $0.32$    &  $0.20$  &  $0.34$  &  $0.11$  \\		
				609.0220    &  $0.22$  &  $-0.15$    &  $0.31$    &  $0.06$  &  $0.28$  &  $-0.08$  \\	
				611.9520    &  $0.23$  &  $-0.16$    &  $0.29$    &  $0.02$ &  $0.28$  &  $-0.10$\\	
				619.9190    &  $0.33$  &  $0.04$     &  $0.29$    &  $0.12$  &  $0.35$  &  $0.07$\\	
				624.3100    &  $0.27$  &  $0.06$     &   --       &  --      &  $0.30$  &  $0.10$\\	
				625.1820    &  $0.28$  &  $-0.06$    &  $0.31$    &  $0.09$  &  $0.30$  &  $-0.03$\\	
				627.4650    &  $0.29$  &  $0.00$    &  $0.30$    &  $0.13$  &  $0.28$  &  $0.00$\\	
				650.4165    &  $0.39$  &  --   &  $0.34$  & --  &   $0.40$  & -- \\				
				\noalign{\vskip 0.6mm}
				\hline\hline
				\noalign{\vskip 1mm}   
				Average    &  $0.27 \pm 0.04$  &  $-0.03 \pm 0.09$    &  $0.30\pm 0.01$    &   $0.11\pm 0.05$  &   $0.30\pm 0.03$  &   $0.01\pm 0.07$\\
			\end{tabular}
		}
	\end{center}
	Note: The average values in the last row do not include the 650.4165\,nm line because it was not used by \citet{Ernandes18}. The errors denote the standard deviation due to the star-to-star abundance variation.
\end{table*}

\clearpage

\begin{table*}
	\begin{center}
		\caption{Abundances of Fe, Na, and Sr determined in the sample of 31 RGB stars in 47~Tuc.
			\label{app-tab:data-list}}
		\begin{tabular}{ccccccccccc}
			\hline\hline
			\noalign{\smallskip}
			\textit{Gaia} Source ID & ID  & $\Teff$ & $\log g$  & $\xi_{\rm micro}$ &  \emph{A}(Fe)  &  [Fe/H] &  \emph{A}(Na)  &  [Na/Fe]  &  \emph{A}(Sr)  &  [Sr/Fe]  \\
		&	&   K     &           &      \,km\,s$^{-1}$         &         &         &         &           &         &           \\
			\hline\noalign{\smallskip}
			
4689640950470780288	&	00251029-7158280	&	4133	&	1.18	&	1.64	&	6.75	&	0.80	&	5.85	&	0.48	&	2.33	&	0.26\\  
4689640774370214528	&	00242489-7159320	&	4235	&	1.32	&	1.55	&	6.77	&	0.78	&	5.88	&	0.49	&	2.17	&	0.08\\
4689624762723599488	&	00234278-7211310	&	4343	&	1.37	&	1.99	&	6.79	&	0.76	&	5.95	&	0.54	&	2.27	&	0.16\\
4689642256141727872	&	00232279-7201038	&	3982	&	0.85	&	1.85	&	6.81	&	0.74	&	5.81	&	0.38	&	2.27	&	0.14\\
4689637407107299840	&	00224044-7208548	&	4114	&	1.09	&	1.58	&	6.75	&	0.80	&	5.86	&	0.49	&	2.27	&	0.20\\
4689643935459451776	&	00235438-7158355	&	4143	&	1.10	&	1.79	&	6.82	&	0.73	&	5.84	&	0.40	&	2.29	&	0.15\\
4689625999695578496	&	00241344-7211263	&	4118	&	1.11	&	1.73	&	6.67	&	0.88	&	5.94	&	0.65	&	2.23	&	0.24\\
4689644730042501760	           &	1062	           &	3962	&	0.87	&	1.50	&	6.77	&	0.78	&	5.80	&	0.41	&	2.26	&	0.17\\
4689651086596023040	           &	4794	           &	4071	&	1.15	&	1.46	&	6.72	&	0.83	&	5.66	&	0.32	&	2.07	&	0.03\\
4689651151009833728	           &	5265	           &	3948	&	0.91	&	1.60	&	6.74	&	0.81	&	5.85	&	0.49	&	2.20	&	0.14\\
4689652117388263296	           &	5968	           &	3959	&	0.93	&	1.45	&	6.71	&	0.84	&	5.54	&	0.21	&	2.15	&	0.12\\
4689625724796264192	           &	6798	           &	4109	&	1.14	&	1.46	&	6.73	&	0.82	&	5.63	&	0.28	&	2.19	&	0.14\\
4689626240193985280	           &	10237	           &	4289	&	1.34	&	1.75	&	6.63	&	0.92	&	5.47	&	0.22	&	2.11	&	0.16\\
4689614970198135296	           &	13396	           &	4245	&	1.34	&	1.44	&	6.67	&	0.88	&	5.57	&	0.28	&	2.25	&	0.26\\
4689638918944031616	           &	20885	           &	4358	&	1.41	&	1.76	&	6.69	&	0.86	&	5.64	&	0.33	&	2.29	&	0.28\\
4689642599739335040	           &	27678	           &	3958	&	0.87	&	1.56	&	6.76	&	0.79	&	5.87	&	0.49	&	2.19	&	0.11\\
4689642256141727872	           &	28956	           &	3982	&	0.85	&	1.86	&	6.77	&	0.78	&	5.75	&	0.36	&	2.18	&	0.09\\
4689830508836705664	           &	29861	           &	4217	&	1.32	&	1.54	&	6.66	&	0.89	&	5.59	&	0.31	&	2.09	&	0.11\\
4689640602571354496	           &	38916	           &	4173	&	1.22	&	1.41	&	6.80	&	0.75	&	5.84	&	0.42	&	2.18	&	0.06\\
4689643630531051392	           &	40394	           &	3951	&	0.82	&	1.69	&	6.57	&	0.98	&	5.73	&	0.54	&	2.30	&	0.41\\
4689623049035687040	           &	s5270	           &	4111	&	1.12	&	1.67	&	6.68	&	0.87	&	5.99	&	0.69	&	2.30	&	0.30\\
4689625690444642560	           &	s12272	           &	4198	&	1.25	&	1.56	&	6.76	&	0.79	&	5.93	&	0.55	&	2.34	&	0.26\\
4689626652510887168	           &	s13795	           &	4262	&	1.34	&	1.77	&	6.70	&	0.85	&	5.56	&	0.24	&	2.15	&	0.13\\
4689638403539699584	           &	s14583	           &	4319	&	1.52	&	1.54	&	6.75	&	0.80	&	5.79	&	0.42	&	2.33	&	0.26\\
4689626893029604096	           &	s17657	           &	4057	&	1.04	&	1.55	&	6.79	&	0.76	&	6.03	&	0.62	&	2.32	&	0.21\\
4689638678425880576	           &	s18623	           &	4323	&	1.47	&	1.43	&	6.84	&	0.71	&	5.95	&	0.49	&	2.27	&	0.11\\
4689638678425891712	           &	s20002	           &	4285	&	1.39	&	1.54	&	6.78	&	0.77	&	5.87	&	0.47	&	2.31	&	0.21\\
4689639473003821440	           &	s23821	           &	4403	&	1.46	&	1.43	&	6.69	&	0.86	&	5.82	&	0.51	&	2.31	&	0.30\\
4689639782230993664	           &	s34847	           &	4145	&	1.20	&	1.55	&	6.70	&	0.85	&	5.65	&	0.33	&	2.28	&	0.26\\
4689640430772703232	           &	s36828	           &	4193	&	1.33	&	1.51	&	6.72	&	0.83	&	5.73	&	0.39	&	2.27	&	0.23\\
4689643630522760960	           &	s41654	           &	4215	&	1.30	&	1.52	&	6.83	&	0.72	&	5.97	&	0.52	&	2.30	&	0.15\\
	
			\hline
		\end{tabular}
	\end{center}
\end{table*}
	
\end{appendix}

\end{document}